\documentclass[aps,prr,reprint,superscriptaddress,numbers,floatfix,colorlinks=true, linkcolor=blue, citecolor=blue, urlcolor=blue, table]{revtex4-2}

\input{packages.sty}
\begin{document}
\definecolor{mypink}{RGB}{255,102,153} 
\definecolor{mygreen}{RGB}{102,204,102} 
\definecolor{mypurple}{RGB}{153,153,204} 
\definecolor{myyellow}{RGB}{204,153,51} 
\definecolor{myblue}{RGB}{102,153,204} 
\definecolor{myorange}{RGB}{255,102,51} 

\captionsetup[figure]{justification=raggedright,singlelinecheck=false}
\newcolumntype{C}[1]{>{\centering\arraybackslash}p{#1}}

\title[Identifying statistical indicators of temporal asymmetry using a data-driven approach
]{Identifying statistical indicators of temporal asymmetry using a data-driven approach
} 

\author{Teresa Dalle Nogare \orcidlink{0009-0002-1983-9079}}
\thanks{Contact author: teresa.dallenogare@sydney.edu.au}
\author{Ben D. Fulcher \orcidlink{0000-0002-3003-4055}}%
\thanks{Contact author: ben.fulcher@sydney.edu.au}
\affiliation{School of Physics, The University of Sydney, Camperdown, NSW 2006, Australia}
\affiliation{Centre for Complex Systems, The University of Sydney, Camperdown, NSW 2006, Australia}

\date{November 20, 2025}

\begin{abstract}
The dynamics of time-reversible systems are statistically indistinguishable when observed forward or backward in time.
A rich literature of statistical methods to distinguish irreversible dynamics from the reversible dynamics of linear, Gaussian systems can provide insights into underlying mechanisms and aid modeling and statistical quantification of time-series data.
But these existing time-reversibility metrics have been developed individually, forming a fragmented body of research that makes it challenging to identify the most effective approaches developed to date, and the most promising new directions for development.
Here we address these issues by systematically evaluating over 6000 time-series summary statistics, derived from across the time-series analysis literature, on their ability to distinguish the time-irreversibility of data simulated from a diverse range of 35 systems.
Our large-scale data-driven comparison highlights the effectiveness of several key families of statistics, including time-asymmetric forms of generalized autocorrelation functions, time-series symbolic sequences, and forecasting-related methods.
All irreversible systems studied here could be accurately distinguished by a well-chosen time-series statistic, but no single statistic could accurately index the statistical form of irreversibility for all irreversible systems.
This challenges the assumption that a given time-reversibility statistic will accurately capture time reversibility in general, and underscores the importance of tailoring statistical approaches to the time-reversal characteristics of a given system.
Our results provide a unified understanding of the key algorithmic structures through which irreversibility can be effectively quantified from data, providing a foundation for connecting patterns in time series to the underlying mechanisms of the systems that generate them.
\end{abstract}

\maketitle 

\section{Introduction}
\label{sec:intro}

The nature of time and the existence of a privileged direction in its flow have long fascinated both philosophers \cite{lopez_review_2024} and scientists, inspiring fundamental questions about the origin of time \cite{rovelli_order_2019}, the neural mechanisms of time perception \cite{buonomano_your_2017}, and the evolution of the universe \cite{hawking_brief_2011}.
The concept of time-reversibility has been studied across several disciplines, including time-series analysis \cite{lawrance_directionality_1991}, dynamical systems theory \cite{lamb_time-reversal_1998}, and non-equilibrium statistical mechanics \cite{peliti_stochastic_2021}.
Its cross-disciplinary importance arises from the ubiquity of time-reversal asymmetric processes in the real world, motivating studies in both abstract deterministic and stochastic systems as well as in the thermodynamics of physical systems.
A deterministic dynamical system is considered reversible if its governing equations remain unchanged under time-reversal \cite{roberts_chaos_1992,lamb_time-reversal_1998} (cf. more general definitions, including for multidimensional systems \cite{devaney_reversible_1976, strogatz_nonlinear_2015}).
When studied in the context of nonequilibrium physical systems, irreversibility acquires a probabilistic interpretation, traditionally expressed in thermodynamics through the Second Law of Thermodynamics as entropy production at the macroscopic scale.
More recently, in stochastic thermodynamics---which has emerged as a theoretical framework that extends concepts from classical thermodynamics (e.g., entropy, heat, work) to smaller-scale systems where thermal fluctuations play a dominant role in the dynamics \cite{seifert_stochastic_2025}---the Kullback--Leibler divergence between the probability distribution of systems trajectories evolving forward and backward in time links irreversibility (i.e. the distinguishability between the process unfolding forward versus backward in time) and entropy production, an indicator of the underlying dissipative mechanisms driving the process \cite{kawai_dissipation_2007, roldan_entropy_2012,roldan_irreversibility_2014}.

This study focuses on the statistical interpretation of time irreversibility (hereafter referred to as `irreversibility'), which refers to the ability to infer the temporal direction of a dynamical process from its observed dynamics.
That is, are the dynamics of some process statistically different from its time-reversed dynamics?
Formally, a discrete-time stationary process $\bm{X} = \{X_t\}_{t\in \mathbb{Z}}$ is said to be \textit{reversible} if, for every time point $t \in \mathbb{Z}$ and every $k \geq 1$, the joint distribution of the sequence $\bm X_t^{(k)}\equiv(X_t, X_{t+1}, \dots, X_{t+k})$ is identical to that of its time-reversed counterpart $\tilde{\bm X}_t^{(k)}\equiv (X_{t+k}, X_{t+k-1}, ..., X_t)$, that is 
\begin{equation}
    \bm X_{t}^{(k)} \stackrel{d}{=} \tilde{\bm X}_t^{(k)},~~~\forall t\in \mathbb{Z} \text{ and } \forall k\geq 1\,,
    \label{eq:rev}
\end{equation}
where the symbol $\stackrel{d}{=}$ denotes that $\bm X_{t}^{(k)}$ and $\tilde{\bm X}_t^{(k)}$ are identically distributed \cite{weiss_time-reversibility_1975,lawrance_directionality_1991}.
In other words, reversibility reflects the time-reversal symmetry of the probabilistic structure of a process \cite{giannakis_time-domain_1994}, such that all of its statistical properties are invariant under time reversal \cite{diks_reversibility_1995}.

In practice, we often seek to perform inference on the reversibility of a generative process from a single finite realization, i.e., a time series $\bm x = (x_1,x_2,...,x_T)$.
In this case, the time-reversed transformed time series $\tilde{\bm x}$ can be constructed by reversing the order of the data points, i.e. $\tilde{x}_t = x_{T-t+1}$ \cite{arola-fernandez_irreversibility_2023, lacasa_time_2015, gonzalez-espinoza_arrow_2020, camassa_temporal_2024}.
Under the assumption of stationarity, the reversibility condition, Eq.~\eqref{eq:rev}, can then be inferred from the time-series data by estimating the joint distributions of $\bm x_{t}^{(k)} = (x_t,x_{t+1},...,x_{t+k})$ and $\tilde{\bm x}_t^{(k)} = (x_{t+k}, x_{t+k-1},...,x_t)$, and quantifying their discrepancy using a suitable statistical distance or divergence measure (e.g., Kullback--Leibler divergence).
That is, the reversibility condition in Eq.~\eqref{eq:rev} can be expressed as
\begin{equation} 
 \bm x_{t}^{(k)} \stackrel{d}{=} \tilde{\bm x}_t^{(k)},~~~\forall t<T-k \text{ and } \forall k<T\,.
 \label{eq:empirical_rev}
\end{equation}

While obtaining accurate estimates of the joint probability distributions $p(\bm x_{t}^{(k)})$ and $p(\tilde{\bm x}_t^{(k)})$ is practicable for low $k$ (e.g., the typical time-delay phase space corresponding to $k = 1$), estimating high-dimensional probability distributions corresponding to larger $k$ quickly becomes impractical for realistic time-series lengths.
Diverse statistical methods have been developed to more tractably do inference on the time-reversibility condition on finite data (i.e., assess violation of the symmetry condition Eq.~\eqref{eq:empirical_rev}) by quantifying changes in time-series properties under time reversal in terms of real-valued statistics or through test statistics that can form the basis of a statistical test for reversibility \cite{cox_statistical_1981,diks_reversibility_1995}.
In this work, we focus on the inference of time reversibility based on a single real-valued summary statistic (or feature) extracted from a time series (i.e., a feature map $f:\mathbb{R}^T \to \mathbb{R}$) that aims to encapsulate the deviation from the time-reversibility condition Eq.~\eqref{eq:empirical_rev}.

The notion of time reversibility has attracted considerable attention in the time-series analysis literature, as it serves as a valuable diagnostic tool in model development and as a means for identifying nonlinearities and non-Gaussianity in the underlying dynamical structure of a system.
A key result established in time-series analysis is that irreversibility reflects departures from linearity and Gaussianity in the underlying process \cite{weiss_time-reversibility_1975}.
This relationship has important implications for model development, since time-irreversible dynamics cannot be generated by canonical linear Gaussian time-series models, thus ruling out this ubiquitous class of dynamics \cite{lawrance_directionality_1991,stone_detecting_1996}.  
Furthermore, reversibility statistics can act as powerful statistical summaries of the dynamical structure of time-series data that can be used for subsequent statistical learning applications like classification problems.
For example, in medicine, reversibility captures differences between healthy controls and patients with epilepsy \cite{van_der_heyden_time_1996, pijn_nonlinear_1997, schindler_ictal_2016, zhang_nonequilibrium_2023}, demonstrating its potential for clinical diagnosis.
Reversibility has also been shown to relate to brain organization and the emergence of wakefulness \cite{camassa_temporal_2024}, while serving as an indicator of abnormalities \cite{costa_broken_2005, guzik_heart_2006, porta_temporal_2008} and nonlinearity \cite{braun_demonstration_1998} in human heartbeat dynamics and hand tremor signals \cite{timmer_characteristics_1993}.
In economics and finance, time series such as business cycles often exhibit asymmetric structures, typically characterized by long, gradual expansions followed by sharp contractions \cite{zumbach_time_2009, chen_testing_2000}, making tests for time reversibility useful for classification \cite{flanagan_irreversibility_2016}.
More recently, irreversibility indicators based on autocorrelation functions \cite{pomeau_symetrie_1982} have revealed scale-dependent time-reversal asymmetry in fully developed turbulence \cite{schmitt_scaling_2023} and identified energy transfer from large to smaller structures in turbulent flows \cite{josserand_turbulence_2017}.

The literature on statistical methods for inferring time-reversal asymmetry from time series is vast, spanning multiple domains and theoretical approaches, and includes a wide range of techniques designed to index time-reversibility from finite time series.
Higher-order cumulants and bi- or polyspectral statistics \cite{brillinger_1967} represent some of the earliest statistical descriptors sensitive to the direction of time, along with asymmetric autocorrelation functions (i.e., forms of autocorrelation that are not invariant under time reversal) \cite{pomeau_symetrie_1982, steinberg_time_1986} and foundational tests based on sample bicovariances, developed both in the time \cite{ramseyj_time_1988, ramseyj_time_1996} and frequency domain \cite{hinich_frequency-domain_1998}.
Other methods are based on measuring asymmetry in the distribution of consecutive temporal intervals with respect to zero, quantified by the third cumulant of the differences $x_{t+\tau} - x_t$ \cite{cox_statistical_1981}, including for processes with long-range dependencies \cite{cox_long-range_1991}, also extended across higher dimensions \cite{casali_multiple_2008}. 
\citet{diks_reversibility_1995} introduced a reversibility test based on the invariance of the distribution of delay vectors $\bm y_n = (x_n,x_{n+\tau}, \dots, x_{n+(m-1)\tau}) \in \mathbb{R}^m$, built from a time series $\bm x$ and across various embedding dimensions $m$ and lags $\tau$, under time reversal.
In two-dimensional embeddings $(x_t, x_{t+\tau})$, the reversibility condition, Eq.~\eqref{eq:empirical_rev}, implies symmetry about the identity line, a property that is harder to visualize in higher dimensions, where it must still hold for all $m$ and $\tau$. 
This phase-space symmetry underlies established irreversibility measures based on time-series increments \cite{tsay_model_1992, porta_time_2006, guzik_heart_2006}, which consider the probabilities of rises and falls in the time series.
Following a similar rationale, \citet{costa_broken_2005} proposed a computational method based on coarse-graining time-series increments at different resolutions, thereby incorporating physical assumptions and extending the quantification of irreversibility across multiple temporal scales.
A nonlinear predictor was also used for irreversibility detection by comparing forecasting performance a fixed number of steps ahead in both the forward and backward dynamics \cite{stone_detecting_1996}.
Other approaches transform time series into alternative representations (such as symbolic sequences or network structures) which are then analyzed using tools from symbolic dynamics \cite{daw_symbolic_2000}, graph theory \cite{lacasa_time_2012, fdonges_testing_2013}, or information theory \cite{kennel_testing_2004}.

Here, we focus on two key limitations of the existing literature on time reversibility: its fragmentation across multiple disciplines, and the tendency for methods to be studied in isolation and evaluated on small, hand-selected sets of systems.
Addressing these limitations would yield several benefits in the literature which include clarifying conceptual relationships across results, highlighting common underlying interpretations rather than isolated outcomes, and enhancing constructive communication across diverse fields.
The breadth of algorithmic contributions to the time-reversibility literature---developed and adopted over many decades and across fields, each with its own terminology, tools, and notation---has contributed to a disjoint literature leading to a practical ambiguity in how to select an appropriate statistic from this wide array of alternatives.
Since these various methods have been proposed and tested individually, there is currently no systematic framework to guide researchers in selecting appropriate approaches from this vast interdisciplinary methodological literature on irreversibility.
In particular, since the strengths and weaknesses of different formulations of time-series statistical structure have seldom been benchmarked or compared to each other in either discrete-time or continuous-time systems \cite{zanin_algorithmic_2021,zanin_algorithmic_2025}, it is not well understood which types of approaches are the most powerful at capturing different types of time-reversibility from data, or what the relative strengths and limitations of each method are across systems containing different types of dynamical structure.
From the viewpoint of method development, we lack a unified context in which to assess whether new and promising methods from the broader time-series analysis field can be effectively adapted to tackle the problem of irreversibility.
Furthermore, approaches in the time-series analysis literature are often validated on a limited range of deterministic systems---most commonly the H\'enon map, logistic map, and Lorenz system---despite theory on time-reversibility covering a wide range of systems that deviate in a range of different ways from linear Gaussian dynamics.
Benchmarking methods on narrow and inconsistent sets of processes can obscure their relative strengths and weaknesses in capturing the diverse ways irreversibility manifests across systems, including different types of temporal asymmetries and varying timescales.

In this work, we aim to address these issues by empirically evaluating a large library of interpretable statistics to identify those most effective for detecting irreversibility and by unifying existing time-series approaches to time-reversibility analysis.
To this end, we systematically compare and analyze the most comprehensive collection of over 6000 time-series statistics for their ability to capture the time-reversal asymmetry of over 35 diverse processes, spanning a range of formulations and dynamical behaviors (in discrete and continuous time, and encompassing varying degrees of linearity and Gaussianity).
We draw on the extensive interdisciplinary library of time-series statistics contained in the highly comparative time-series analysis package, \textit{hctsa} \cite{fulcher_highly_2013, fulcher_hctsa_2017}, in which each feature is derived from interpretable time-series theory.
This connection to theory facilitates an understanding of the types of temporal structures (and ways of quantifying them) that diverse scientists have developed to date which are relevant to distinguishing reversibility.
The highly comparative, data-driven approach to methodological comparison has yielded new understanding and guided the selection of novel time-series methods for a wide range of applications---identifying features most associated with altered intrinsic brain dynamics in epileptic patients \cite{yang_macroscale_2024}, discovering novel heart rate variability metrics \cite{letzkus_heart_2023}, and demonstrating the utility of some established time irreversibility features in distinguishing zebra finch songs \cite{paul_behavioral_2021}.
It has also facilitated the development of new theory for challenging empirical problems like tracking the proximity to criticality in near-critical systems with unknown noise levels \cite{harris_tracking_2024}.
Here we apply this approach to the problem of time reversibility for the first time to develop a unified understanding of the relative strengths and weaknesses of different types of time-series analysis methods across different types of dynamical processes, allowing us to synthesize the array of disparate prior work on this topic.

The paper is organized as follows.
In Sec.~\ref{sec:methods}, we introduce our highly comparative time-series analysis approach for detecting irreversibility.
We describe how we generate synthetic data from various simulated processes and outline the procedure for extracting time-series features.
The results of our analysis are presented in Sec.~\ref{sec:results}.
We differentiate between features that are insensitive to the direction of time across all simulated processes (Sec.~\ref{subsec:zero_features}) before focusing on the subset of high-performing statistical indicators of irreversibility (Sec.~\ref{subsec:well_features}).
We provide a detailed interpretation of the identified families of well-performing features, with particular emphasis on generalized autocorrelation functions, symbolic measures, and forecasting-based statistics.
After identifying and interpreting the top-performing features, we compare their effectiveness in distinguishing between reversible and irreversible dynamics across different processes in Sec.~\ref{subsec:spectrum_rev}.
Finally, in Sec.~\ref{sec:discussion}, we provide concluding remarks, discuss strengths and limitations of our data-driven methodology, and draw future directions for the problem of irreversibility detection from time-series data.

\section{Methods}
\label{sec:methods}
\begin{figure*}[htbp]
    \centering

        {\includegraphics[width=\textwidth]{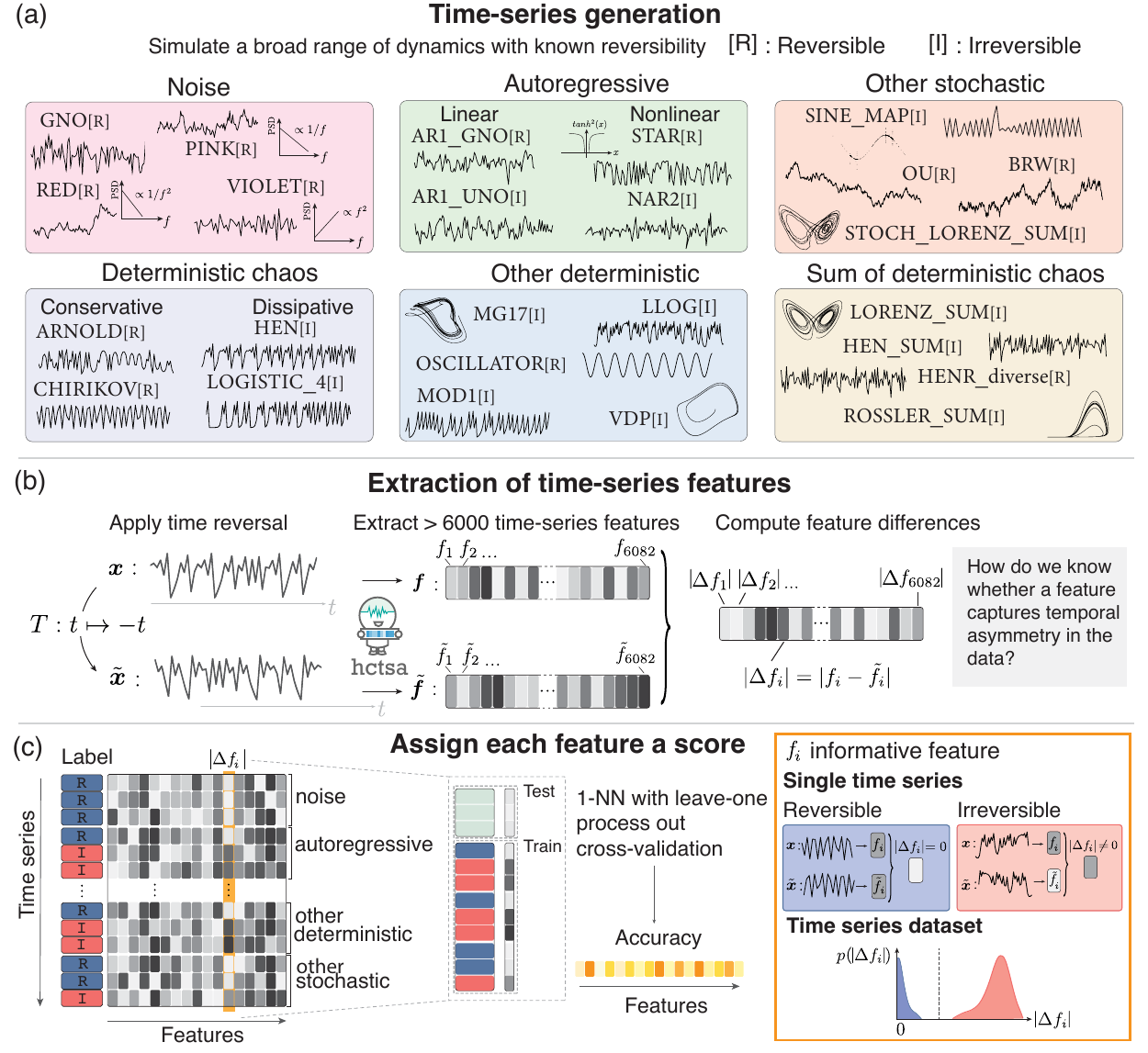}
        \phantomsubcaption\label{subfig:methods-a}
        
        \phantomsubcaption\label{subfig:methods-b}
        
        \phantomsubcaption\label{subfig:methods-c}}
    
    \caption{
    \textbf{Schematic of our data-driven approach to identifying high-performing time-series statistics that can accurately index irreversibility from time-series data.}
    \textbf{(a) Time series generation:} To include multiple and diverse sources of irreversibility, we simulated $5000$-sample time series from a comprehensive library of 35 discrete-time and continuous-time processes with known reversibility properties.
    Each panel shows selected illustrative time series segments from a given process class, with a [R] indicating `reversible' and a [I] indicating `irreversible', placed near the process label according to the reversibility specified in Table~\ref{tab:processes}.
    We use a consistent color coding throughout the paper to identify these families of simulated processes.
    \textbf{(b) Extraction of time-series features:} For each time series $\bm x$, we constructed its time-reversed counterpart $\tilde{\bm x}$ by reversing the order of the data points, as per prior work \cite{arola-fernandez_irreversibility_2023,lacasa_time_2015,gonzalez-espinoza_arrow_2020,camassa_temporal_2024}.
    To find time-series properties that are sensitive to irreversibility, we systematically computed a large set of $>6000$ interpretable time-series features implemented in the \textit{hctsa} library \cite{fulcher_hctsa_2017}.
    In this way, each time series was summarized by a set of real numbers, $f_1, \dots, f_{6082}$, that encode a broad range of its statistical properties.
    As an ansatz for the ability of each time-series feature $f_i$ to index a difference in statistical properties between the original time series $\bm x$ and the time-reversed time series $\tilde{\bm x}$, we quantified the absolute difference between the feature value computed on $\bm x$ and $\tilde{\bm x}$, as $|\Delta f_i| = |f_i - \tilde{f}_i|$.
    \textbf{(c) Assign each feature a score:} To extract time-series properties that are sensitive to the reversibility of processes, each feature $f_i$ was scored based on the discriminative power of its $|\Delta f_i|$ in classifying time series generated by reversible [R] and irreversible [I] processes using a 1-nearest-neighbor classifier (1-NN) with leave-one-out cross-validation; the classification accuracy served as the feature's score. We expect that an informative feature $f_i$, associated with a high classification accuracy, displays $\Delta f_i \approx 0$ for reversible processes and large deviations $|\Delta f_i| > 0$ for irreversible ones, yielding well-separated distributions across the dataset.
    }
    \label{fig:method}
\end{figure*}

This paper takes a highly comparative approach to understanding types of time-series analysis approaches that are useful for inferring irreversibility from time-series data.
Our data-driven methodology is illustrated schematically in Fig.~\ref{fig:method} and can be summarized in three main steps:
(i) generation of time-series data by simulating stochastic processes and dynamical systems with known time-reversal properties (depicted in Fig.~\ref{subfig:methods-a}). 
The set of 35 simulated processes is introduced in Sec.~\ref{subsec:model_systems} and further elaborated in Sec.~\ref{subsec:time_series_generation}, where implementation details are also provided (see Appendix~\ref{app:models});
(ii) evaluation of a comprehensive set of time-series features to identify those most sensitive to temporal asymmetry (depicted in \cref{subfig:methods-b,subfig:methods-c}).
The procedure for feature extraction and the scoring framework used to highlight the most informative features are described in Sec.~\ref{subsec:feature_extraction}; and
(iii) interpretation of top-performing features to understand how they quantify different types of temporal patterns associated with irreversibility (see Sec.~\ref{sec:results}).
 
\subsection{Model systems}
\label{subsec:model_systems}

A core component of our data-driven method is a way to meaningfully score each time-series feature on its ability to distinguish time series generated by reversible versus irreversible processes.
For such a scoring process to be useful, the set of systems we test on should be sufficiently diverse to capture as wide a range of dynamical signatures of irreversibility that are exhibited by different classes of dynamical processes.
By contrast, if our set of systems were too small or specific, features could score well by capturing idiosyncratic properties of specific processes, instead of the broader characteristic of irreversibility.
To this end, we generated a collection of univariate, stationary time series simulated from 35 reversible and irreversible systems (we simulated 100 time series from each system), yielding a dataset of 3500 time series for studying irreversibility that contains the most comprehensive coverage of processes to date (to our knowledge).
Our collection includes a broad range of both stochastic and deterministic, discrete- and continuous-time processes, subdividing processes that share similar dynamics into families, as summarized in Table~\ref{tab:processes} and illustrated in Fig.~\ref{subfig:methods-a}, where each panel corresponds to a family of simulated processes consistently color-coded throughout the paper.
In selecting processes to incorporate in this work, we aimed to include coverage of as many diverse dynamical behaviors as possible, while generally aligning with the parameter settings used in previous studies to ensure a consistent reference for evaluating the reversibility of the simulated systems.
In labeling each system as either `reversible' or `irreversible', we could, in some cases, rely on prior knowledge of the process (as per the reversible i.i.d. noise processes), but in other cases relied on existing literature on irreversibility and systems that have been used to assess time-reversibility metrics to base the assessment.

Given that the dynamics of real-world systems are often modeled in the existing literature using ordinary or stochastic differential equations, we included continuous-time processes in our analysis alongside discrete-time formulations.
When treated computationally, however, numerical solutions to such systems inevitably involve approximations or truncations \cite{hoover_time_1998}, as further discussed in Sec.~\ref{subsec:time_series_generation}.
Despite the added complexity of numerically handling continuous-time systems compared to their discrete-time counterparts, including representative examples of these systems is essential to capture dynamics with varying time-scale dependencies and to evaluate the robustness of individual statistics. 
Overall, we included 35 diverse systems: 15 reversible systems and 20 irreversible systems.
The motivation for including each type of dynamics, along with references to prior work used to assess their reversibility, is presented below, while details of the simulated systems are provided in Appendix~\ref{app:models}.

\setlength{\tabcolsep}{0pt}

\begin{table*}
\caption{\label{tab:processes} Summary of simulated processes grouped into six families, with a color coding that is used throughout the paper.
Each process is labeled with an abbreviation for ease of reference.
The discrete- or continuous-time nature of the simulated process is indicated by `D' for discrete and `C' for continuous.
The reversibility of each system is indicated by `R' for reversible and `I' for irreversible (assigned based on either referenced prior studies or from ground truth knowledge of the process; cf. Appendix~\ref{app:models} for details for specific systems).
}
\rowcolors{2}{gray!10}{white}
\begin{tabular}{ C{0.2\linewidth} C{0.4\linewidth} C{0.2\linewidth} C{0.1\linewidth} C{0.1\linewidth} }

\toprule
\textbf{Family of processes} & \textbf{Process} & \textbf{Label} & \textbf{Discrete/ Continuous} & \textbf{Reversibility}\\ \hline
\cellcolor{mypink!20}\textbf{Noise} & Gaussian noise \cite{diks_reversibility_1995} & \textsc{GNO} & D & R \\
\cellcolor{mypink!20}& Uniform noise \cite{diks_reversibility_1995} & \textsc{UNO} & D & R\\ 
\cellcolor{mypink!20}& Pink noise \cite{gonzalez-espinoza_arrow_2020}& \textsc{PINK} & D & R \\
\cellcolor{mypink!20}& Red noise \cite{gonzalez-espinoza_arrow_2020}& \textsc{RED} & D & R \\
\cellcolor{mypink!20}& Violet noise \cite{gonzalez-espinoza_arrow_2020}& \textsc{VIOLET} & D & R\\\hline
\cellcolor{mygreen!20}\textbf{Autoregressive models} & AR(1) with i.i.d. Gaussian noise\cite{weiss_time-reversibility_1975, diks_reversibility_1995} & \textsc{AR1\_GNO} & D & R\\
\cellcolor{mygreen!20}& Static transformation of AR(1) with i.i.d. Gaussian noise\cite{weiss_time-reversibility_1975, diks_reversibility_1995} & \textsc{STAR} & D & R \\
\cellcolor{mygreen!20}& AR(1) with i.i.d. uniform noise \cite{diks_reversibility_1995} & \textsc{AR1\_UNO} & D & I \\
\cellcolor{mygreen!20}& ARMA(1,1) with i.i.d. uniform noise & \textsc{ARMA11\_UNO} & D &  I \\
\cellcolor{mygreen!20}& AR(3) with noise sampled from a Gamma distribution & \textsc{AR3\_GAMMA} & D &  I \\
\cellcolor{mygreen!20}& SETAR(2;1,1,1)\footnote{Self-Exciting Threshold Autoregressive Model}\cite{mittnik_higher-order_1999}& \textsc{SETAR1} & D & I\\
\cellcolor{mygreen!20}& SETAR(2;2,2,2)\cite{martinez_detection_2018} & \textsc{SETAR2} & D & I \\
\cellcolor{mygreen!20}& Nonlinear AR(2) with non-Gaussian noise\cite{martinez_detection_2018} & \textsc{NAR2} & D & I\\\hline
\cellcolor{mypurple!20}\textbf{Deterministic chaotic maps} & Arnold's Cat map \cite{lacasa_time_2012} & \textsc{ARNOLD} & D & R\\
\cellcolor{mypurple!20}& Chirikov standard map \cite{roberts_chaos_1992} & \textsc{CHIRIKOV} & D & R\\
\cellcolor{mypurple!20}& H\'enon map\cite{diks_reversibility_1995, pomeau_symetrie_1982} & \textsc{HEN} & D & I\\
\cellcolor{mypurple!20}& Quadratic map\cite{kennel_testing_2004} & \textsc{QUAD} & D & I \\
\cellcolor{mypurple!20}& Logistic map ($r=4$)\cite{stone_detecting_1996, lacasa_time_2012, arola-fernandez_irreversibility_2023} & \textsc{LOGISTIC\_4} & D & I \\
\cellcolor{mypurple!20}& Logistic map ($r=3.8284$)\cite{arola-fernandez_irreversibility_2023} & \textsc{LOGISTIC\_38} & D & I \\ \hline 
\cellcolor{myyellow!20}\textbf{Sum of deterministic chaotic maps/flows}
& Sum of H\'enon maps, one forward and the other reversed\cite{diks_reversibility_1995} & \textsc{HENR\_DIVERSE} & D & R\\
\cellcolor{myyellow!20}& Sum of Hénon map and its reversed & \textsc{HENR\_SAME} &  D & R\\
\cellcolor{myyellow!20}& Sum of a quadratic maps, one forward and the other reversed\cite{kennel_testing_2004} & \textsc{QUAD\_RSUM} & D & R\\
\cellcolor{myyellow!20}& Sum of H\'enon maps\cite{diks_reversibility_1995} & \textsc{HEN\_SUM} & D & I \\
\cellcolor{myyellow!20} & Sum of Lorenz \cite{pomeau_symetrie_1982,stone_detecting_1996,martinez_detection_2018} & \textsc{LORENZ\_SUM} & C & I\\
\cellcolor{myyellow!20}& Sum of R\"ossler \cite{martinez_detection_2018}& \textsc{ROSSLER\_SUM} & C & I \\ \hline
\cellcolor{myblue!20}\textbf{Other deterministic models} & Linear model with logistic noise source\cite{diks_reversibility_1995} & \textsc{LLOG} & D & I \\
\cellcolor{myblue!20} & Modulus-1 model\cite{diks_reversibility_1995} & \textsc{MOD1} & D & I \\ 
\cellcolor{myblue!20}& Van der Pol oscillator \cite{pomeau_symetrie_1982}& \textsc{VDP} & C & I \\
\cellcolor{myblue!20} & Oscillator & \textsc{OSCILLATOR} & C & R \\
\cellcolor{myblue!20}& Mackey--Glass \cite{fdonges_testing_2013} & \textsc{MG17} & C & I\\ \hline
\cellcolor{myorange!20}\textbf{Other stochastic processes} 
& Stochastic sine map & \textsc{SINE\_MAP} & D & I \\
\cellcolor{myorange!20}& Ornstein--Uhlenbeck \cite{lacasa_time_2012} & \textsc{OU} & C & R \\
\cellcolor{myorange!20}& Bounded random walk & \textsc{BRW} & C & R \\
\cellcolor{myorange!20}& Stochastic Van der Pol oscillator & \textsc{STOCH\_VDP} & C & I \\
\cellcolor{myorange!20}& Sum of stochastic Lorenz & \textsc{STOCH\_LORENZ\_SUM} & C & I \\
\bottomrule
\end{tabular}
\end{table*}

\paragraph*{Noise.}
We included both uncorrelated (i.e., white noise) and long-range correlated (i.e., colored noise) stochastic processes under a common class labeled `\textit{noise}' in Table~\ref{tab:processes} for convenience, as a class of processes that lack any deterministic component.
We simulated time-series realizations of independent and identically distributed (i.i.d.) noise processes drawn from both a Gaussian (labeled \textsc{GNO}) and a uniform distribution (labeled \textsc{UNO}).
Time series consist of uncorrelated datapoints, drawn from a fixed distribution, leading to invariant joint distributions under time reversal, indicating that time series are statistically reversible. 
Given the ground-truth knowledge of their reversible nature, these processes serve as canonical examples in empirical studies of time symmetry \cite{diks_reversibility_1995,gonzalez-espinoza_arrow_2020}.
In addition, we included three colored noise processes, generated as time series with a given power spectrum: pink noise (labeled \textsc{PINK}), red noise (labeled \textsc{RED}), and violet noise (labeled \textsc{VIOLET}).
Since the power spectrum is insensitive to temporal irreversibility due to its inherently time-reversal symmetric construction \cite{steinberg_time_1986}, colored noise processes are reversible.

\paragraph*{Autoregressive processes.}
We analyzed time series generated from a set of autoregressive-moving-average (ARMA) models with different linearity and Gaussianity properties, focusing on two reversible and six irreversible processes, as summarized in Table~\ref{tab:processes}.
The theoretical foundation of reversibility in linear stochastic processes---established in the seminal work of \citet{weiss_time-reversibility_1975}, which established a link between Gaussianity, linearity, and time-reversibility---makes these processes canonical for studying numerical methods for indexing reversibility.
As representatives of reversible processes, we considered a first-order autoregressive process with Gaussian noise (labeled \textsc{AR1\_GNO}) and, given that the reversibility of a linear stochastic Gaussian process is preserved under any point transformation \cite{weiss_time-reversibility_1975}, a second process obtained by applying a static nonlinear transformation to the first (labeled \textsc{STAR})\cite{diks_reversibility_1995}. 
Although nonlinearities or non-Gaussianities alone do not guarantee the irreversibility of linear stochastic processes \cite{weiss_time-reversibility_1975,cox_statistical_1981}, we accounted for modifications of standard linear Gaussian processes whose irreversibility has been established in prior studies \cite{diks_reversibility_1995, mittnik_higher-order_1999, martinez_detection_2018}.
We examined three irreversible autoregressive models with non-Gaussian noise distributions: a first-order autoregressive process AR(1) with uniform noise (labeled \textsc{AR1\_UNO}) \cite{weiss_time-reversibility_1975,diks_reversibility_1995}, a first-order autoregressive and moving average process ARMA(1,1) (labeled \textsc{ARMA11\_UNO})\cite{weiss_time-reversibility_1975} and a more complex form of linear dynamics which accounts for longer memory dependencies combined with noise from a Gamma distribution, implemented through a third-order autoregressive process, AR(3) (labeled \textsc{AR3\_GAMMA}).
Next, we simulated three nonlinear variants of autoregressive models by introducing nonlinearities either through regime switching \cite{tong_threshold_1980} or through equations involving squared terms and other functional forms.
Specifically, we included two variants of the Self-Exciting Threshold Autoregressive, SETAR, model previously identified as irreversible (labeled \textsc{SETAR1} and \textsc{SETAR2}) \cite{mittnik_higher-order_1999,martinez_detection_2018}, and a nonlinear second-order autoregressive process, AR(2) (labeled \textsc{NAR2})\cite{martinez_detection_2018}.

\paragraph*{Deterministic chaotic maps.}
Motivated by prior work investigating the relationship between reversibility and chaos \cite{roberts_chaos_1992}, we included various examples of chaotic dynamics \cite{pomeau_symetrie_1982, diks_reversibility_1995,lacasa_time_2012,arola-fernandez_irreversibility_2023, stone_detecting_1996, kennel_testing_2004} (see Table~\ref{tab:processes}).
Despite the conceptual distinction between reversible and conservative dynamics \cite{roberts_chaos_1992}, we included two representative discrete-time examples of chaotic maps that are both measure-preserving and reversible, namely the Arnold's Cat map (labeled \textsc{ARNOLD}) \cite{lacasa_time_2012} the Chirikov standard map (labeled \textsc{CHIRIKOV}) \cite{roberts_chaos_1992}, as well as four that are dissipative and irreversible: the H\'enon map (labeled \textsc{HEN}) \cite{diks_reversibility_1995,pomeau_symetrie_1982}, the quadratic map (labeled \textsc{QUAD}) \cite{kennel_testing_2004}, and the logistic map in two chaotic regimes (labeled \textsc{LOGISTIC\_4} and \textsc{LOGISTIC\_38}, respectively) \cite{arola-fernandez_irreversibility_2023,stone_detecting_1996, lacasa_time_2012}.

\paragraph*{Sum of deterministic chaotic maps/flows.}
In Table~\ref{tab:processes} we included linear combinations of realizations of discrete-time chaotic dynamics as an interesting extension to investigate how reversibility changes between individual realizations and their linear combinations (e.g., from the irreversible nature of the H\'enon map to the reversible nature of the linear combination between a realization and its time-reversed counterpart \cite{diks_reversibility_1995}).
Specifically, we generated diverse linear combinations derived from the H\'enon and quadratic maps, including both: (i) combinations of forward realizations that preserve the irreversible dynamics characteristic of these processes (labeled \textsc{HEN\_SUM}) \cite{diks_reversibility_1995}; and (ii) combinations of a forward realization with time-reversed versions, which yield reversible dynamics (labeled \textsc{HENR\_DIVERSE} when the reversed series is a distinct realization, and \textsc{HENR\_SAME} when the reversed series corresponds to the same realization, and \textsc{QUAD\_RSUM} where the reversal is generated from a diverse realization) \cite{diks_reversibility_1995,kennel_testing_2004}.

To enrich our dataset with continuous-time processes, we included linear combinations of the $x$, $y$, and $z$ components of the dynamics of two canonical examples of continuous-time chaotic flows: the Lorenz \cite{pomeau_symetrie_1982,stone_detecting_1996,martinez_detection_2018} (labeled \textsc{LORENZ\_SUM}) and R\"ossler (labeled \textsc{ROSSLER\_SUM}) \cite{martinez_detection_2018} systems (see Table~\ref{tab:processes}).
We labeled these linear combinations as `irreversible' as they contain components that prior work has demonstrated to be irreversible; that is any irreversible component yields a violation to the time-reversibility condition.

\paragraph*{Other deterministic models.}
To incorporate other nonlinear and deterministic non-chaotic processes in our analysis, we implemented two models whose irreversible character has been previously studied by \citet{diks_reversibility_1995}: a linear model with deterministic noise (labeled \textsc{LLOG}) and a modulus-1 process (labeled \textsc{MOD1}) (see Table~\ref{tab:processes}).

We included two kinds of continuous-time dynamic behaviors that extend beyond first-order ordinary differential equations: those described by second-order differential equations and those governed by delay differential equations.
As representative of dynamics described by second-order differential equations, we included a reversible linearized oscillator (labeled \textsc{OSCILLATOR}), a prototypical example often used to intuitively introduce the concept of reversibility, and an irreversible Van der Pol oscillator (labeled \textsc{VDP}) \cite{pomeau_symetrie_1982}.
Physical systems characterized by time delays are particularly interesting from the perspective of reversibility, as their dynamics explicitly depend on past values; accordingly, our dataset simulated time series from the nonlinear time-delayed Mackey–Glass equations (labeled \textsc{MG17}) \cite{fdonges_testing_2013}.

\paragraph*{Other stochastic processes.}
Our dataset also encompasses stationary random walks, as well as chaotic and oscillatory dynamics corrupted by i.i.d. noise (see Table~\ref{tab:processes}).
A discrete-time process for which we analyze the irreversibility consists of a nonlinear deterministic map perturbed by additive non-Gaussian noise (labeled \textsc{SINE\_MAP}).

In continuous-time, we included two examples of reversible random walks, namely the widely studied time-reversible Ornstein--Uhlenbeck process (labeled \textsc{OU}) \cite{lacasa_time_2012} and a bounded random walk (labeled \textsc{BRW}).
To study the effect of stochastic fluctuations on the reversibility of dynamical systems, we examine two irreversible systems---the sum of the $x$, $y$, and $z$ components of the Lorenz system (labeled \textsc{STOCH\_LORENZ\_SUM}), and the Van der Pol oscillator (labeled \textsc{STOCH\_VDP})---perturbed by noise, which we label as `irreversible' based on the their underlying deterministic dynamics.

\subsection{Time-series generation}
\label{subsec:time_series_generation}
To evaluate the performance of each time-series feature at distinguishing the diverse set of 15 reversible and 20 irreversible processes described above, we simulated a total of 100 time series from each process, yielding a combined dataset containing $3500$ simulated time series.
Continuous-time stochastic processes were simulated by numerical integration using the Euler--Maruyama method \cite{kloeden_numerical_1992} with an integration time step $\Delta t = 10^{-2}$\,s.
Continuous-time deterministic systems were simulated by numerical integration using the Runge--Kutta--Fehlberg (RKF45) method \cite{fehlberg_low-order_1969} with a default integration step $\Delta t = 10^{-2}$\,s (with two exceptions: $\Delta t = 1$\,s for the Mackey--Glass system (\textsc{MG17}), which is sufficient to resolve the dynamics relative to the characteristic timescale set by the delay parameter, and $\Delta t = 10^{-3}$\,s for the bounded random walk (\textsc{BRW}), chosen to balance numerical accuracy and computational efficiency).
Time series were simulated using the Python language (version 3.12.4), except for colored i.i.d. noise series for which we used MATLAB 2020a.

For our main analyses, we selected a time-series length of $T = 5000$ samples for all systems, striking a compromise between being sufficiently long (to enable complex methods to capture signatures of irreversibility), while remaining realistic for many empirical applications involving finite data.
To avoid transient effects of the initial condition, for discrete-time processes we initially simulated 10,000 samples and then discarded the first 5000 samples, resulting in the $T = 5000$ sample time series analyzed here.
For the continuous-time case, we first simulated each continuous-time process over a sufficiently long time interval (accounting for the fact that the number of samples is inversely proportional to the integration step) to discard the initial transient.
We confirmed that this transient duration exceeded the longest characteristic timescale (estimated as the first zero-crossing of the autocorrelation function) for all systems in our simulations, indicating that it is sufficient to eliminate any dynamics related to the initial condition of any given simulation, which are specified for each model in Appendix~\ref{app:models}.
To ensure that all processes were sampled on a relatively comparable timescale relative to the dynamics of the process (and avoid substantial under-sampling or over-sampling, which can bias many time-series statistics operating over relatively short temporal lags), we downsampled time series generated from continuous-time processes by reducing its sampling rate of an integer factor $m$, chosen as the temporal lag at which the autocorrelation function decays to $1/e$.
While this approach of estimating a `correlation length' is more naturally suited to stochastic processes, using this common processing heuristic for all continuous-time processes allowed us to effectively tackle the problem of setting a sampling rate that similarly resolves the relevant dynamical correlations for both deterministic and stochastic systems.
After downsampling, we retained the first $T = 5000$ samples of each series.
The sensitivity of results to this time-series length $T$ is analyzed briefly for selected time-series features in Appendix~\ref{app:robustness}.

\subsection{Feature extraction}
\label{subsec:feature_extraction}
From our diverse library of simulated time series from reversible and irreversible processes, we aimed to identify and characterize the types of time-series analysis methods that can most effectively distinguish the two types of dynamics.
To this end, we adopted a highly comparative, data-driven framework, which involved systematically evaluating $>7000$ candidate time-series features from the \textit{hctsa} toolbox (version 1.09) \cite{fulcher_highly_2013, fulcher_hctsa_2017} using MATLAB 2020a.
Each feature corresponds to a single real-valued statistical summary of a $T$-sampled long time-series property, $f:\mathbb{R}^T \to \mathbb{R}$ ($T = 5000$ here), and is typically computed after z-scoring the time series prior to feature extraction.
Examples of features implemented in \textit{hctsa} include statistics of the distribution of time-series values (including outlier measures), diverse measures of linear and nonlinear temporal correlation structure, as well as a range of complexity and dimensionality metrics \cite{fulcher_hctsa_2017}.
The time-series feature-extraction process is illustrated in Fig.~\ref{subfig:methods-b}.
In particular, for each simulated time series $\bm{x} = (x_1,\dots,x_T)$, we constructed its time-reversed counterpart, $\tilde {\bm x} = (\tilde x_1,\dots, \tilde x_T)$ by reversing the order of the data points, $\tilde{x}_t = x_{T+1-t}$, as per prior work \cite{arola-fernandez_irreversibility_2023,lacasa_time_2015,gonzalez-espinoza_arrow_2020,camassa_temporal_2024}.
We then extracted the same set of $7796$ features from both $\bm x$ and $\tilde{\bm x}$, yielding two feature vectors: $\bm f =\{f_i\}_{i\in \mathbb{N}}$ and $\tilde{\bm f}=\{\tilde{f}_i\}_{i\in \mathbb{N}}$, in which each entry of each feature vector corresponds to the output of some interpretable time-series analysis algorithm.
After removing features that could not be successfully evaluated across the entire dataset (i.e., all forward and all time-reversed time series), we obtained a consistent set of $6082$ time-series features used throughout the remainder of this work.

Given the set of candidate features, our next goal was to identify those which are most effective for detecting reversibility.
To this end, we adopted the \textit{ansatz} that the top-performing statistics can be identified via a test statistic defined for each feature $f_i$ (with $i=1,....,6082$) as the absolute difference between the values computed on the original and time-reversed time series:
\begin{equation}
    |\Delta f_i| \equiv |f_i - \tilde f_i|\,.
    \label{eq:deltaf}
\end{equation}
Intuitively, $|\Delta f_i|$ captures the discrepancy in a given statistical property $f_i$ between the original and time-reversed series, and can thereby capture a source of statistical irreversibility.
The construction in Eq.~\eqref{eq:deltaf} is motivated by the similar form used in previous test statistics for reversibility (e.g., the `TR test' statistic\cite{ramseyj_time_1996}, or correlation-based measures\cite{pomeau_symetrie_1982}).
Note that $|\Delta f_i| = 0$ corresponds to $f_i = \tilde{f_i}$ and thus time-reversal invariance, where a given time-series property $f_i$ is evaluated identically for both $\bm{x}$ and $\tilde{\bm x}$.
Since we expect $|\Delta f_i|\approx 0$ for reversible processes, discriminating statistics $f_i$ for irreversibility should be able to capture deviations from the condition $\Delta f_i = 0$ in the case of irreversible dynamics.
Significant deviations from zero reflect the ability of the statistic to capture the statistical change in a time series under time reversal.
Note that, since the sign of $\Delta f_i$ is unrelated to the goal of assessing deviations from the $\Delta f_i \approx 0$ condition, taking the absolute value as $|\Delta f_i|$ provides us with a desired test statistic for indexing a deviation from reversibility.

As illustrated in Fig.~\ref{subfig:methods-c}, we next aimed to develop a scoring statistic to evaluate features $f_i$ for which $|\Delta f_i|$ is highly discriminative of time series generated from reversible versus irreversible processes.
To this end, we used the performance of a 1-nearest neighbor (1-NN) classifier in the space of each feature $f_i$, evaluated using a leave-one-process-out cross-validation strategy.
For a given feature $f_i$, this approach consists of matching each time series to that with the closest $f_i$ value (1-NN), while excluding all time series generated from the same process (leave-one-process-out).
The resulting performance score for $f_i$ is then computed as the rate at which the matches were correctly assigned to time series of the same (`reversible'/`irreversible') label.
This leave-one-process-out exclusion prevented overly optimistic performance estimates that could arise from a feature simply capturing characteristic properties of a given process (rather than more general statistical structure related to irreversibility).
Note that, to avoid bias from the arbitrary ordering of systems, when multiple points had the identical distance to the query point for 1-NN, the matching neighbor was selected randomly.

As explained above, since $|\Delta f_i| \approx 0$ for reversible processes, in practice, high-performing features are those for which $|\Delta f_i|$ deviates substantially from zero for a wide range of irreversible processes, resulting in irreversible time series tending to match (via leave-one-process-out 1-NN) to other irreversible processes.
We thus found this 1-NN heuristic scoring procedure to be a suitable empirical metric with which to identify time-series features that were discriminative of irreversibility.

All code required to generate the time series and reproduce the analyses presented in this paper is openly available on GitHub at \cite{code_github}.
The supporting data for the simulations presented in this article are available from Zenodo \cite{data_paper}.
\section{Results}
\label{sec:results}
In this work, we aim to develop a unified understanding of the most successful types of existing time-series analysis approaches for inferring the reversibility or irreversibility of a process from a finite time series realization $\bm x$.
In this section and throughout the remainder of this work, $\bm x$ refers to z-scored time series data.
As depicted in Fig.~\ref{fig:method}, our data-driven approach involves comparing over 6000 time-series features on their ability to distinguish time series generated from a comprehensive range of 35 reversible and irreversible processes, using a derived statistic capturing the difference between the feature computed from the forward ($f$) versus reversed ($\tilde{f}$) time series, as $|\Delta f|$, Eq.~\eqref{eq:deltaf}.
In Sec.~\ref{subsec:zero_features}, we first characterize time-series features that are invariant to a time-reversal transformation (i.e., yielding $\Delta f \approx 0$ for all time series tested) and interpret them with respect to their underlying time-symmetric constructions.
Next, in Sec.~\ref{subsec:well_features}, we characterize the range of existing and novel time-series analysis methods that are most successful at quantifying irreversibility, and explain the algorithmic structures that underlie their strong performance.
Finally, in Sec.~\ref{subsec:spectrum_rev}, we demonstrate that there is no optimal summary statistic for capturing irreversibility from time-series data in general; rather, all features we tested have strengths and weaknesses that depend on how the irreversibility of a given process manifests.

\subsection{Time-reversal invariant time-series features}
\label{subsec:zero_features}

We first aimed to characterize the type of time-series features that are invariant under the time-reversal transformation of a time series.
Such time-reversal-invariant features are interesting to characterize initially because they provide a foundation from which we can later understand features that can distinguish the direction of time.
Well-known time-reversal-invariant time-series statistics include properties of the distribution of time-series values (e.g., mean), which disregard the sequential ordering in the data (and so are insensitive to any permutation, including time reversal), and simple two-point linear autocorrelation statistics for some lag $\tau$, as $\langle x_t\, x_{t+\tau} \rangle$, where $\langle \cdot \rangle$ denotes the temporal average over the time series (and statistics of the related Fourier power spectrum, cf. the Wiener--Khinchin theorem \cite{cohen_time_1995}) \cite{pomeau_symetrie_1982, lawrance_directionality_1991, steinberg_time_1986}.
Note that the time-reversal symmetry of the two-point linear autocorrelation statistic can be determined straightforwardly by considering the equivalence of the terms in the average that makes up $\langle x_t \, x_{t+\tau} \rangle$ or, for a time series of finite length $T$, by explicitly analyzing a time-reversal transformation (as $t \mapsto T-t-\tau + 1$, cf. Appendix~\ref{app:symm_ac} for details).

To generate a candidate list of time-reversal-invariant features from our set of $6082$ tested time-series statistics, we identified those features $f_i$ that gave $\Delta f_i \approx 0$ across all tested time series, corresponding to giving identical outputs (approximately, within numerical error) when applied to each original time series $\bm x$ and its time-reversed version $\tilde{\bm{x}}$.
To account for numerical error, we used the heuristic criterion $|\overline{\Delta f}| < 5\varepsilon$, where $\overline \cdot$ denotes the average over all time series and $\varepsilon = 2.2 \times 10^{-16}$ is the floating-point precision in Python.
This yielded a set of $1414$ time-reversal-insensitive time-series features (see Supplemental Material Table~S1 for full list).

This set of time-reversal-insensitive features includes expected distributional properties (e.g., mean, median, standard deviation, and quantiles) and two-point linear autocorrelation statistics (and related properties of the power spectrum), as well as a range of conceptually related statistics with time-symmetric constructions.
These include symmetric generalizations of autocorrelation functions (which we refer to as generalized autocorrelations, introduced later in Sec.~\ref{subsec:generalized_ac}), measures of symmetric patterns of time-series rises and falls (analyzed in more detail in Sec.~\ref{subsec:symbolic}), some information-theoretic statistics such as auto-mutual information \cite{kraskov_estimating_2004} (which is time-reversal symmetric for the same reason as the two-point linear autocorrelation statistic), and some statistics of the degree distribution derived from the undirected horizontal visibility graph \cite{luque_horizontal_2009} (note that subsequent developments have introduced directionality to network edges to capture irreversibility \cite{lacasa_time_2012}).

Our results recapitulate prior literature on statistics that capture a diverse range of time-series properties but are trivially insensitive to irreversibility due to an invariance under the time-reversal transformation---either by insensitivity to temporal ordering (as distributional measures), or via a time-symmetric construction.
These time-symmetric constructions will be explored in greater detail, and contrasted with time-asymmetric constructions in Sec.~\ref{subsec:well_features}.

\subsection{Time-series statistics of reversibility}
\label{subsec:well_features}

\begin{figure*}[htpb]
    \centering
    {\includegraphics[width=\linewidth]{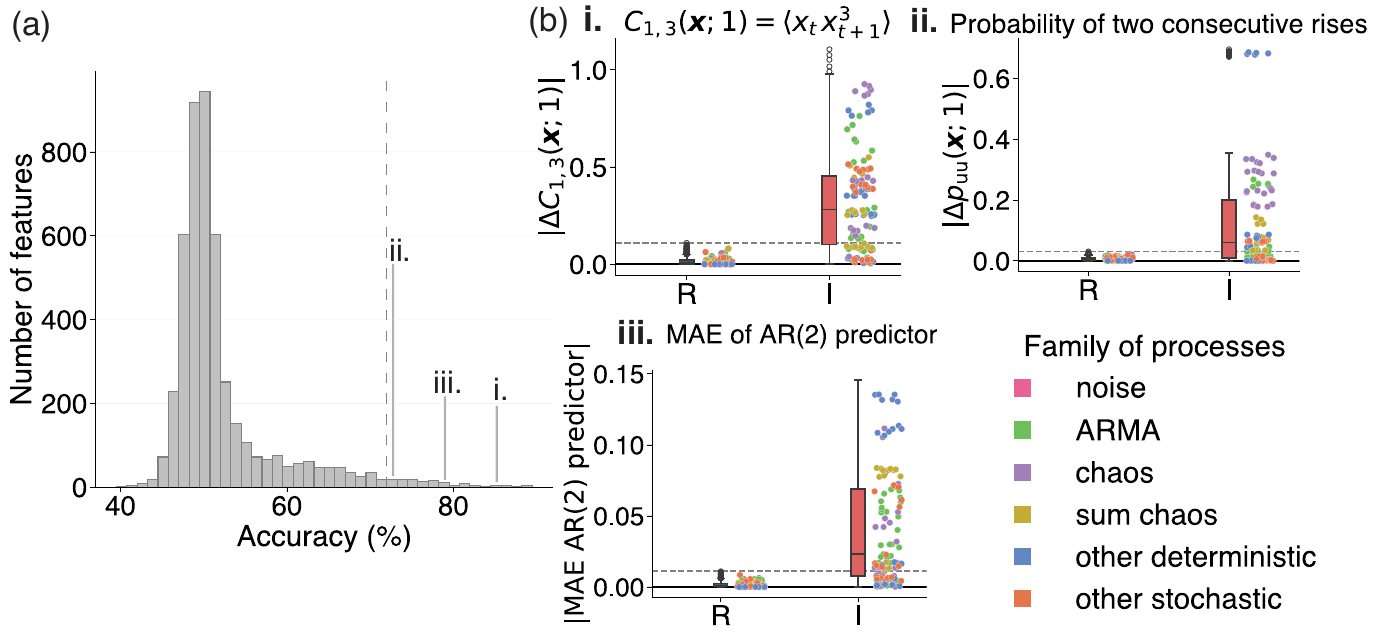}
        \phantomsubcaption\label{subfig:dist-a}
        
        \phantomsubcaption\label{subfig:dist-b}}
        
    \caption{
    \textbf{Identifying high-performing and interpretable time-series features for detecting irreversibility through large-scale empirical testing of thousands of features on 35 reversible and irreversible processes}.
    \textbf{(a)} Distribution of cross-validated classification accuracy (of distinguishing reversible from irreversible processes) across $4668$ features (excluding $1414$ that are insensitive to reversibility, cf. Sec.~\ref{subsec:zero_features}).
    Each feature $f_i$ was assessed independently on its ability to distinguish time series generated from reversible versus irreversible processes via the absolute difference between its value computed on the time series and its time-reversed counterpart, as $|\Delta f_i|$ (Eq.~\eqref{eq:deltaf}).
    There is a tail in the accuracy distribution, pointing to a subset of high-performing features; here we focus on the $127$ features with accuracies exceeding $72\%$, annotated as a dashed gray vertical line (see Supplementary Material Table S3 for a full list).
    \textbf{(b)} Distributions of $|\Delta f|$ across all 3500 time series, separated between 1500 reversible (blue) and 2000 irreversible (red) time series, are shown as box plots for three selected top-performing features:
    \textbf{i.} the fourth-order cross-moment (an example of generalized autocorrelation-based feature) $C_{1,3}(\bm x;1)=\langle x_t\, x_{t+1}^3\rangle$;
    \textbf{ii.} the symbolic motif feature, $p_{\text{uu}}(\bm x)$, which calculates the probability of two consecutive rises in a time series, and
    \textbf{iii.} the mean absolute error (MAE) of 1-step-ahead predictions made by a second-order autoregressive (AR) model.
    Next to each boxplot is a raincloud plot showing the $|\Delta f|$ value for all time series and colored according to process families (defined in Table~\ref{tab:processes}), with random horizontal scatter to aid visualization.
    Horizontal black lines indicate the zero baseline, while gray dashed lines delimit the range of values of the statistics computed from time series generated by reversible processes.
    The generalized autocorrelation (i.), symbolic motif (ii.), and MAE of a 1-step-ahead AR(2) prediction (iii.) are annotated in the distribution in Fig.~\ref{subfig:dist-a}.
    These three features were chosen as demonstrative examples of broader families of easily interpretable time-reversal-sensitive time-series features, generalized autocorrelations (Sec.~\ref{subsec:generalized_ac}), symbolic motif probabilities (Sec.~\ref{subsec:symbolic}), and forecasting-based measures (Sec.~\ref{subsec:forecasting}), which are explained in detail in the main text.
    Their behavior mirrors that of all top-performing features, exhibiting $|\Delta f| \approx 0$ for reversible processes but deviating substantially from zero for many irreversible processes.
    \label{fig:distribution}}
\end{figure*}

Having characterized the types of features that are, by construction, invariant under time reversal, we next aim to characterize time-series statistics that are highly effective at quantifying the time-reversal asymmetry present in time series generated from irreversible processes.
Specifically, after removing the $1414$ time-reversal-invariant features characterized above, we aimed to understand the relative performance of the remaining $4668$ features (listed in Supplementary Material Table~S2) and, in particular, understand the best-performing types of time-series summary statistics for indexing time reversibility.
Recall that we assessed each feature's ability to capture irreversibility from time-series data according to its accuracy at classifying time series simulated from 35 reversible and irreversible processes (in a leave-one-process-out cross-validation design, cf. Sec.~\ref{subsec:feature_extraction}).
We expected effective irreversibility statistics to quantify temporally asymmetric time-series properties that capture the non-Gaussianity and/or nonlinearity of irreversible processes.
This includes traditional measures of nonlinearity \cite{schreiber_discrimination_1997}, statistics based on sample bicovariances (the `TR test' statistic) \cite{ramseyj_time_1996} or higher-order moments \cite{cox_long-range_1991}, asymmetric autocorrelation functions \cite{pomeau_symetrie_1982}, and statistics of symbolic strings computed via symbolization rules applied to real-valued time-series data \cite{daw_symbolic_2000}.

The distribution of accuracy values across the $4668$ candidate time-series features is shown in Fig.~\ref{subfig:dist-a}.
We see that the majority of these features are not sensitive to irreversibility, with a large peak around chance accuracy (50\%).
This suggests that the majority of features in the time-series analysis literature (as represented by the \textit{hctsa} library \cite{fulcher_hctsa_2017}) are mainly focused on the statistics of linear, Gaussian processes and are thus insensitive to time-reversal asymmetry.
The distribution of accuracies also exhibits a striking right-skewed tail, indicating the presence of a subset of high-scoring features.
To understand the types of algorithmic approaches that were most effective at indexing irreversibility, we focused on a subset of 127 features with an accuracy exceeding 72\% (shown as a dashed vertical line in Fig.~\ref{subfig:dist-a}).
This threshold was selected to encompass a sufficiently large set of features for analysis, while retaining a focus on those with the highest accuracy.
This set of high-performing features (listed in Supplementary Material Table~S3) is analyzed in the remainder of this paper and encompasses a surprising range of time-series analysis algorithms, including types of statistics that have previously been used to study reversibility (e.g., estimators of higher-order moments and symbolic representations of sequential rise and fall sequences) as well as a range of novel approaches that have (to our knowledge) not previously been used to study reversibility, including the goodness of fit of a range of time-series forecasting models. 

To better understand the behavior of highest-performing features, we highlight three representative features that illustrate the different manifestations of irreversibility revealed by our data-driven analysis: (i) the fourth-order cross-moment $\langle x_t \,x_{t+1}^3 \rangle$ (85\% accuracy, in Fig.~\ref{subfig:dist-b}(i)) which is an example of what we refer to as a generalized autocorrelation feature (defined in Sec.~\ref{subsec:generalized_ac}); (ii) the probability of two consecutive rises in the time series, $p_{\text{uu}}(\bm x)$ (73\% accuracy, in Fig.~\ref{subfig:dist-b}(ii)); and (iii) the mean absolute error (MAE) of 1-step-ahead predictions made by a second-order autoregressive (AR) model (79\% accuracy, in Fig.~\ref{subfig:dist-b}(iii)).
These three plots show the distributions of absolute feature-value differences $|\Delta f|$ which are, as expected, tightly centered around $|\Delta f| \approx 0$ for reversible processes, reflecting the equivalence of the statistics of these processes under time-reversal.
In contrast, distributions of $|\Delta f|$ display a substantial deviation $|\Delta f| > 0$ for many of the irreversible processes (reflecting an asymmetry between the feature computed on the forward versus time-reversed time series).
Compared to reversible processes (with $\Delta f \approx 0$), time series generated by irreversible processes exhibit a far greater heterogeneity in $|\Delta f|$ values across different types of systems, as indicated by the coloring by process in raincloud points in Fig.~\ref{subfig:dist-b}.
In particular, for all high-performing features, there were some irreversible processes for which the source of irreversibility was not captured by the real-valued statistic (i.e., for which $|\Delta f| \approx 0$).
This suggests that, among the set of analyzed features, single statistical index of reversibility will be sensitive to only a subset of ways in which irreversibility can manifest in the data, a concept that will be explored in detail in Sec.~\ref{subsec:spectrum_rev}.

Most high-performing statistics were members of the three conceptual families of time-series analysis methods depicted in Fig.~\ref{subfig:dist-b}:
generalized autocorrelation features (defined in Sec.~\ref{subsec:generalized_ac}, like $\langle x_t\, x_{t+1}^3 \rangle$, in Fig.~\ref{subfig:dist-b}(i));
statistics of rise/fall patterns (like $p_{\text{uu}}(\bm x)$, in Fig.~\ref{subfig:dist-b}(ii)); and forecasting-based measures (like the MAE of 1-step-ahead predictions made by a second-order autoregressive model, in Fig.~\ref{subfig:dist-b}(iii)).
Accordingly, these three families are explored in depth in dedicated sections Sec.~\ref{subsec:generalized_ac}, Sec.~\ref{subsec:symbolic}, and Sec.~\ref{subsec:forecasting}.
In addition to these three families of reversibility statistics, a range of miscellaneous time-series statistics were also contained in the high-performing set.
This includes a set of twelve `stick-angle' statistics that compute measures of the angles between successive time-series points that have the same sign (after $z$-scoring).
These statistics were introduced into \textit{hctsa} \cite{fulcher_highly_2013} and, although they were not designed to capture reversibility, they are highly sensitive to temporal asymmetries in irreversible processes (with accuracies as high as 89\%) through their ability to quantify a directional bias in the distribution of angles computed on the positive (or negative) part of the time series generated from irreversible processes.
Another group of 27 high-performing statistics are derived from simulations of a `walker' process.
Of these, 22 rely on dynamical rules based on the time series, which can index temporal asymmetry through a difference in walker dynamics when driven forward through time versus backward in time (with accuracies up to 89\%). 
The remaining five are based on extreme-value statistics \cite{altmann_reactions_2006}, which (loosely) capture irreversibility by comparing the temporal spacing between large deviations from the mean between the forward and reversed series (up to 79\%).

In summary, our data-driven evaluation of thousands of time-series features revealed the diversity of existing time-series statistics that can successfully capture temporal asymmetry in a range of irreversible processes.
Different types of time-series methods are able to index different summary statistics that capture the difference between the joint distributions (see Eq.~\eqref{eq:empirical_rev}) across a wide range of tested irreversible processes, with three families of approaches---generalized autocorrelation functions, symbolic features, and forecasting-based statistics---encompassing the most powerful families of methods in our test.
We discuss these families in detail in the following sections.
Note that, although our main analyses focus on a single time-series length, $T = 5000$ samples, additional tests on a selected group of top-performing statistics show that results using this time-series length are broadly representative of those which would be obtained for shorter time series.
In particular, we found the forth-order cross-moment feature $\langle x_t \,x_{t+1}^3 \rangle$, the symbolic feature $p_{\text{uu}}(\bm x)$, and the mean absolute error of an AR(2) predictor, a forecasting-based measure, to be quite robust to changes in the time-series length over the range of 10 to 5000 samples (see Fig.~\ref{fig:appC-robustness}).
We observe that, with around 100 samples, the $\Delta f$ appears to have largely converged across all three statistics and all five processes. 
These preliminary results suggest that comparable outcomes can be expected with sample sizes as small as approximately 100.

\subsubsection{Generalized autocorrelation features}
\label{subsec:generalized_ac}

In Sec.~\ref{subsec:zero_features}, we noted the invariance of the two-point linear autocorrelation statistic $\langle x_t\, x_{t+\tau} \rangle$ under time reversal (as well as related statistics of the Fourier power spectrum) since they are defined as averages over products of time-series values separated by a time lag $\tau$, which is preserved under time reversal.
Generalized forms of autocorrelation extend the standard two-point linear autocorrelation by capturing more complex dependencies between points in a time series.
As defined here, these measures encompass a wide range of operations that, under the assumption of stationarity, capture different types of dependencies between local patterns in a time series, including temporal averages of products of data points and linear combinations of functions of those points.
Here, we focus on two classes of features within the family of autocorrelation-based statistics, constructed from different functional forms of time-averages:
(i) as products of time-series data points raised to different powers (e.g., $\langle x_t \, x_{t+\tau}^2\rangle$), or higher-order correlations with unequally spaced time-series points, such as three-point (or multi-point) autocorrelations (e.g., $\langle x_t \, x_{t+1}\, x_{t+3}\rangle$); and
(ii) as normalized linear combinations of temporal averages of a function $f$ applied to time-series points (e.g., $\langle f(x_t^\alpha\,x_{t+\tau}^\beta)\rangle - \langle f(x_t^\alpha) \rangle\, \langle f(x_{t+\tau}^\beta)\rangle$).
In our experiments, 24 features derived from this family were in the top-performing set described in Sec.~\ref{subsec:well_features} above.

The first group of twelve high-performing features contained features of the form 
\begin{equation}
    C_{\alpha,\beta,\gamma,...}(\bm x;\tau_1,\tau_2,...)=\langle f(x_t^\alpha\, x_{t+\tau_1}^\beta\, x_{t+\tau_2}^\gamma\,...) \rangle \,,
    \label{eq:gac_1}
\end{equation}
for some transformation function $f(\cdot)$, being either the two-point form $\langle x_t^\alpha\,x_{t+\tau}^\beta\rangle$ with unequal exponents $\alpha \neq \beta$ (or with an absolute value as $\langle |x_t^\alpha\,x_{t+\tau}^\beta|\rangle$), or with multiple time lags, such as the three-point form $\langle x_t^\alpha\,x_{t+\tau_1}^\beta\,x_{t+\tau_2}^\gamma \rangle$.
Notably, computing the difference between statistics of this family for the original and time-reversed series effectively captures, among other measures, higher-order moments (i.e., the skewness in the first difference $x_{t + \tau} - x_t$), which have been effectively employed in prior work to assess irreversibility \cite{cox_statistical_1981}.
Specific examples include $\langle x_t \, x_{t + 1}^3 \rangle$ (85\% accuracy), $\langle x_t \, x_{t + 2}^3 \rangle$ (79\%), $\langle x_t \, x_{t + 3}^3 \rangle$ (75\%), and the three-point statistic $\langle x_t^2 \, x_{t+1} \, x_{t+3} \rangle$ (75\%).
Interestingly, features that used an absolute value (to our knowledge not studied previously) exhibited higher performance, including $\langle |x_t^3\, x_{t+1}|\rangle$ (87\% accuracy) and $\langle |x_t^2\, x_{t+1}|\rangle$ (87\%).

A second group of twelve high-performing generalized autocorrelation features were constructed according to the `generalized linear self-correlation function' formulation of \citet{duarte_queiros_yet_2007}.
This function is given by
\begin{equation}
C^\prime_{\alpha,\beta}(\bm x; \tau) \equiv \frac{\langle |x_t|^\alpha \, |x_{t+\tau}|^\beta\rangle - \langle |x_t|^\alpha\rangle \langle |x_{t+\tau}|^\beta\rangle}{\sqrt{\langle x_t^{2\alpha}\rangle -\langle |x_t|^\alpha\rangle^2}\sqrt{\langle x_{t+\tau}^{2\beta}\rangle -\langle |x_{t+\tau}|^\beta\rangle^2}}\,,
\label{eqn:glscf}
\end{equation}
for various values of the exponents $\alpha$, $\beta$, and the time-lag $\tau$.
This functional form was originally introduced in the study of financial markets, where it was designed to capture multiscale dependencies in traded volume \cite{duarte_queiros_yet_2007}.
To the best of our knowledge, the formulation in Eq.~\eqref{eqn:glscf} has not previously been used to study irreversibility.
Nevertheless, our comparative analysis highlighted twelve statistics of this general form, with $\tau = 1$ and various combinations of $\alpha$ and $\beta$ (particularly $\alpha = 1, 2$ and $\beta = 2, 5, 10$), including the highest-performing statistic from this group $C^\prime_{1,2}(\bm x; 1)$ (with 88\% accuracy).

\begin{figure*}[htbp]
    \centering
    {
    \includegraphics[width=0.8\linewidth]{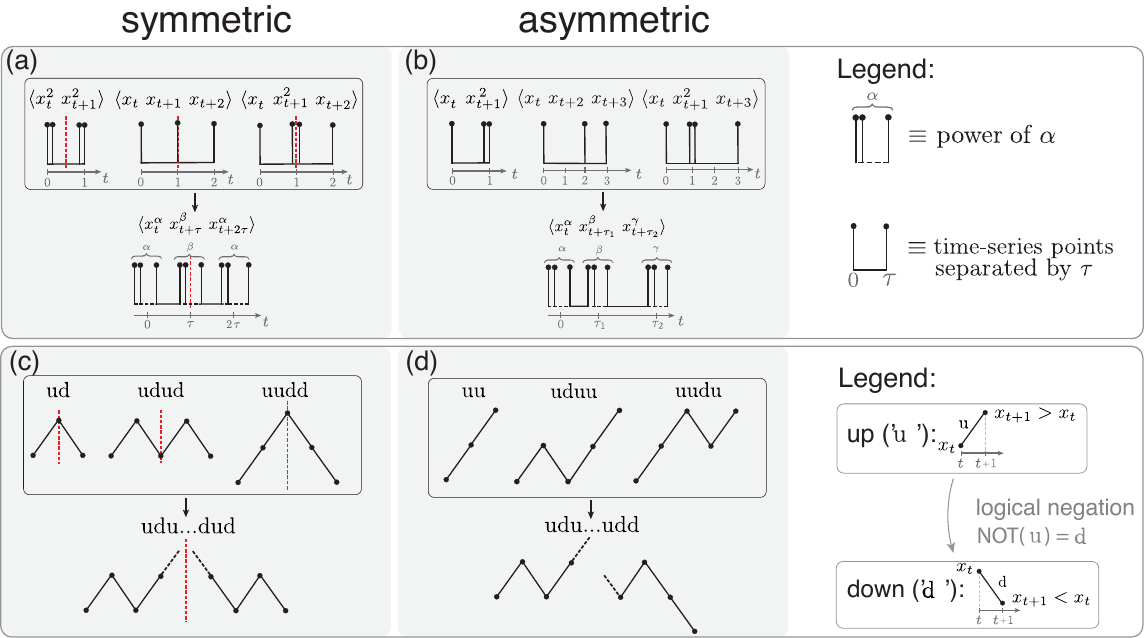}
        \phantomsubcaption\label{subfig:symm-a}
        
        \phantomsubcaption\label{subfig:symm-b}
        
        \phantomsubcaption\label{subfig:symm-c}
        
        \phantomsubcaption\label{subfig:symm-d}
        }

    \caption{
    \textbf{Time-series features with time-symmetric constructions are invariant to time reversal, while time-asymmetric constructions can be powerful indices of time-reversibility.}
    We illustrate this concept with respect to two families of time-series statistics: (a), (b): generalized autocorrelation functions; and (c), (d): the frequency of patterns of consecutive rises (`up': u) and falls (`down': d) in a symbolized transformation of the time series.
    \textbf{(a)} We depict three examples of generalized autocorrelation functions with time-symmetric constructions, visualized diagrammatically using a comb-like representation introduced here.
    In this representation, each vertical segment represents a time-series value at a specific time $t$, the spacing between groups of segments reflects the temporal lag between consecutive feature terms and the number of segments in each group represents the exponent, that is, the number of times a given time-series point contributes to the statistic.
    All of these features are symmetric in time with respect to their midpoint, indicated with a vertical dashed red line.
    \textbf{(b)} Three generalized autocorrelation functions with time-asymmetric constructions are depicted, with the asymmetry arising from the use of different exponents (or non-equally spaced temporal lags).
    \textbf{(c)} Three example time-symmetric sequences of successive rises (`up': u) and falls (`down': d) are depicted diagrammatically.
    The symmetry of these diagrams about their midpoint, depicted as a red dashed line, corresponds to the second half being the mirror image of the logic negation of the first half (indicated by the NOT operation, which transforms `u' to `d' and vice-versa).
    \textbf{(d)} Three examples of time-reversal asymmetric symbolic patterns (which are therefore candidates for indexing irreversibility) are depicted.
    }
    \label{fig:symmetric-asymmetric}
\end{figure*}

Having summarized a set of generalized autocorrelation statistics, we now aim to examine why certain generalized autocorrelation measures are insensitive to time reversal (like the two-point linear autocorrelations $\langle x_t\, x_{t+\tau}\rangle$ and others characterized in Sec.~\ref{subsec:zero_features} above), while others rank among the highest-performing statistics in our empirical tests.
To this end, we focus on the family of statistics given in Eq.~\eqref{eq:gac_1} and examine the distinction between generalized autocorrelations with symmetric versus asymmetric constructions.
This distinction provides a useful framework in which to interpret how these families of statistics can efficiently capture relevant differences between the joint distributions computed from the original time series $\bm x$ ($p({\bm x}_t^{(k)})$) and time-reversed counterpart $\tilde{\bm x}$ ($p(\tilde{\bm x}_t^{(k)})$) (see Eq.~\eqref{eq:empirical_rev}) via a real-valued summary, in Eq.~\eqref{eq:deltaf}.
Illustrative examples of symmetric and asymmetric forms of generalized autocorrelation statistics are shown in Figs~\ref{subfig:symm-a} and \ref{subfig:symm-b}, where we introduce a comb-like representation of the functions of patterns around a given point $t$.
This diagrammatic representation visualizes products of time-series points in the temporal average: each data point is shown as a segment, with the number of segments in a group indicating its exponent in the statistic and the horizontal separation between groups corresponding to the temporal lag $\tau$, as illustrated in the legend of the upper panels of Fig.~\ref{fig:symmetric-asymmetric}.
As will be elaborated through this section, the comb-like visualization emphasizes the temporal symmetry with respect to the midpoint of these functions in constructions with trivial time-reversal symmetry.

Consider a generalized expression of the simple linear two-point autocorrelation computed from a $T$-sample time series $\bm x$, given by
\begin{equation}
C_{\alpha,\beta}(\bm{x};\tau) \equiv \langle x_t^\alpha \, x_{t+\tau}^\beta\rangle \,,
\label{eq:two-point}
\end{equation}
for positive integer exponents $\alpha$ and $\beta$, and time-lag $\tau$.
In this simple case, the construction of $C_{\alpha,\beta}(\bm{x};\tau)$ is asymmetric with respect to time-reversal if $\alpha \neq \beta$, while it reduces to a symmetric form when $\alpha = \beta$ (where symmetric forms are equivalent under time-reversal).
It is straightforward to express $C_{\alpha,\beta}(\tilde{\bm{x}};\tau)$ in terms of the original series $\bm x$ via the transformation $t \mapsto T-t-\tau+1$ (see Appendix~\ref{app:asymm_2ac} for details), yielding the following expression for the absolute difference $|\Delta C_{\alpha,\beta}(\bm{x};\tau)|$:
\begin{equation}
    \begin{aligned}
         |\Delta C_{\alpha,\beta}(\bm{x};\tau)| &= |C_{\alpha,\beta}(\bm{x};\tau) - C_{\alpha,\beta}(\tilde{\bm{x}};\tau)|\,, \\
         &=|\langle x_t^\alpha \, x_{t+\tau}^\beta\rangle - \langle x_{t+\tau}^\alpha \, x_t^\beta\rangle|\,.
    \end{aligned} 
    \label{eq:C_twoPoint}
\end{equation}
Symmetric constructions with $\alpha = \beta$ (where the linear autocorrelation, with $\alpha = \beta = 1$, is a special case, see Sec.~\ref{subsec:zero_features}) are trivial in the sense that $\Delta C_{\alpha,\alpha} = 0$, independent of the data $\bm{x}$.
By contrast, assigning unequal weights to present and future points, with $\alpha \neq \beta$, yields a difference $|\Delta C_{\alpha,\beta}(\bm{x};\tau)|$ which is non-zero in general.
Our experiments confirm the ability to use asymmetric constructions of $C_{\alpha,\beta}$, via $|\Delta C_{\alpha,\beta}(\bm{x};\tau)|$, as a powerful class of statistics for distinguishing irreversible processes (for which we can have $|\Delta C_{\alpha,\beta}(\bm{x};\tau)| > 0$) from reversible processes (for which $|\Delta C_{\alpha,\beta}(\bm{x};\tau)| \approx 0$).

The distinction between symmetric and asymmetric forms of three-point (and, more generally, multi-point) statistics depends on both the choice of weights assigned to each data point and the temporal lags separating successive points (see Appendix~\ref{app:asymm_3ac} for a general expression of the absolute difference in asymmetric three-point statistics).
This distinction can be visualized through the comb-like representations in Figs~\ref{subfig:symm-a} and~\ref{subfig:symm-b}, which depict a general formulation of three-point statistics and highlight the contrast between their symmetric and asymmetric constructions.
Examples of symmetric generalized autocorrelation measures are shown in Fig.~\ref{subfig:symm-a}, illustrating a general symmetric construction of a three-point correlation statistic of the form $\langle x_t^\alpha\,x_{t+\tau}^\beta\, x_{t+2\tau}^\alpha\rangle$ that encompasses and generalizes the symmetric structure of three example statistics: $\langle x_t^2\, x_{t+1}^2\rangle$, $\langle x_t\, x_{t+1}\, x_{t+2}\rangle$, $\langle x_t\, x_{t+1}^2\, x_{t+2}\rangle$.
This general construction highlights two key characteristics: (i) the time points are equally spaced, and (ii) the exponents on the left and right match (while the middle term may be raised to any power).
In contrast, Fig.~\ref{subfig:symm-b} shows examples of associated asymmetric constructions, where asymmetry arises from modifying exponents (e.g., $\langle x_t\, x_{t+1}^2 \rangle$), from unequal spacing between successive pairs of data points (e.g., $\langle x_t \, x_{t+2}\, x_{t+3}\rangle$), or a combination of both (e.g., $\langle x_t \, x_{t+1}^2\, x_{t+3}\rangle$), which is expressed in a general asymmetric form of three-point correlation $\langle x_t^\alpha\,x_{t+\tau_1}^\beta \, x_{t+\tau_2}^\gamma\rangle$.
The same reasoning behind symmetric and asymmetric constructions extends to the generalized autocorrelation $C'_{\alpha,\beta}(\bm x;\tau)$ (Eq.~\eqref{eqn:glscf}) through the choice of the positive integer exponents $\alpha$ and $\beta$.
In particular, our results confirm that the symmetric constructions of $C'_{\alpha,\alpha}(\bm x;\tau)$ are trivially insensitive to detecting irreversibility from time series, whereas asymmetric forms with $\beta \neq \alpha$ can reach high classification accuracies.

The best-performing features in the family of generalized autocorrelations align closely with asymmetric correlation-based statistics proposed in foundational work on time irreversibility, while extending previously studied higher-order moments that also achieve strong performance.
For instance, \citet{pomeau_symetrie_1982} introduced cubic two-point correlation measures such as $\psi^{\prime\prime}(\tau) = \langle x_t^3\,x_{t+\tau} - x_t\, x_{t+\tau}^3\rangle\,$, and \citet{ramseyj_time_1996} proposed the well-known `TR test' statistic for measuring asymmetry in business cycles, $\hat{\!\gamma}_{2,1}(\tau) = \langle x_t^2\, x_{t-\tau}\rangle - \langle x_t\, x_{t-\tau}^2 \rangle\,$.
However, in our empirical tests, the best-performing TR test reached a maximum accuracy of only 68\% (with $\tau = 2$).
By contrast, we found that extending these classical measures through additional transformations not previously studied---such as taking absolute values---led to substantial boosts in performance, e.g., $\langle |x_t^3\, x_{t+1}|\rangle$ and $\langle |x_t^2\, x_{t+1}|\rangle$, both reached 87\% accuracy.
We also examined a widely used nonlinearity measure introduced by \citet{schreiber_discrimination_1997}, $t^\mathrm{rev}(\tau) = \langle (x_t - x_{t-\tau})^3 \rangle\,$, which is a commonly used statistic for time reversal (and practical proxy for nonlinearity) in the physics-based nonlinear time-series analysis literature \cite{cox_statistical_1981}.
Despite its widespread use in nonlinear dynamics, the classification accuracy of $|\Delta t^{rev}|$ remained below 71\% in our tests, across five tested values of the temporal lag ($\tau = 1, 2, \dots 5$).
The similar overall behavior of the TR statistic $\hat{\!\gamma}_{2,1}(\tau)$ and $t^\mathrm{rev}(\tau)$ can be explained by the fact that they convey nearly the same information under the assumptions of stationarity and ergodicity, 
\begin{equation}
    t^{rev}(\tau) = -3\, \hat{\!\gamma}_{2,1}(\tau)\,,
    \label{eq:similar_sch_rr}
\end{equation}
as it can be observed by expanding the cube in the $ t^{rev}(\tau)$ statistic.
\citet{chen_testing_2000} showed the same calculation using ensemble averages, thus considering that testing reversibility through $\mathbb{E}[x_t^2\,x_{t-\tau}] = \mathbb{E}[x_t\, x_{t-\tau}^2]$ is equivalent to testing whether the third moment of the increment series $y_t = x_t-x_{t-\tau}$ vanishes.
This correspondence is noteworthy because it explicitly connects the TR statistic, originally defined for testing reversibility, with the symmetry of the underlying process distribution, while Eq.~\eqref{eq:similar_sch_rr} provides a practical formulation for evaluating this relationship using empirical data.

In this section, we explained how a wide range of asymmetric constructions of generalized forms of autocorrelation are able to index time-reversal asymmetry in time-series data, including through two-point statistics with unequal exponents (e.g., $\langle x_t^2\, x_{t+1}\rangle$), higher-order statistics through the additional freedom introduced by time lags (e.g., $\langle x_t\, x_{t+2}\, x_{t+3}\rangle$), using additional operations like taking magnitudes (e.g., $\langle |x_t^3\, x_{t+1}|\rangle$), and through novel functional forms involving time-averages like the financial statistic $C^\prime_{\alpha,\beta}(\bm x;\tau)$ (which achieved the highest accuracy, 88\%, but has not previously been used to index reversibility) \cite{duarte_queiros_yet_2007}.
This broad family of statistics were among the most successful at quantifying reversibility in our empirical tests, recapitulating existing reversibility statistics from the literature, as well as identifying new forms, such as $\langle x_t^2 \, x_{t+1} \, x_{t+3} \rangle$, which generalize earlier constructions by computing, for example, absolute values of the product of time-series data-points \cite{pomeau_symetrie_1982,ramseyj_time_1988,schreiber_discrimination_1997}.

\subsubsection{Symbolic features}
\label{subsec:symbolic}

The second key family of high-performing time-series statistics that we explore in detail is the group of `symbolic features', which are both well-established for indexing reversibility and straightforward to analyze.
These symbolic features involve transforming a time series $\bm x$ into a sequence of symbols (from a finite alphabet) using a set of coarse-graining rules \cite{daw_symbolic_2000,kennel_testing_2004,martinez_detection_2018}.
Prior work has explored several symbolic approaches, including coarse-graining the time series based on value thresholds \cite{daw_symbolic_2000}, encoding rules that capture relationships between successive time-series values \cite{costa_broken_2005}, and an analysis of the distribution of all possible symbol permutations of fixed length---known as permutation patterns (or ordinal patterns) \cite{bandt_permutation_2002, keller_ordinal_2014,zanin_ordinal_2021}.
Among the set of symbolizing transformations included in \textit{hctsa}, the most effective transformation for capturing irreversibility was the mapping to a two-letter alphabet based on incremental rises and falls in a time series.
This transformation maps each pair of consecutive time-series values to a binary symbol \{u,d\} (`up' and `down') that encodes either a rise ($x_{t+1} - x_t > 0$) or fall ($x_{t+1} - x_t < 0$) in the time series $\bm x$, with the symbol `u' (up) denoting a rise and `d' (down) denoting a fall.
From the resulting symbolic sequence, simple statistics can be derived that capture different temporal structures, such as the length of the longest consecutive run of a given symbol or the probability of occurrence of a given sequence.
A simple example is the probability of the two-letter sequence `ud', $p_{\text{ud}}(\bm x)$: the probability that the time series $\bm x$ increases in one time step ($x_{t+1} > x_t$) and then decreases in the subsequent time step ($x_{t+2} < x_{t+1}$).
Although some symbolic features ranked among the top performers in our tests, their accuracies---generally below 74\%---were lower than that of other features (which reached accuracies as high as 89\%).
Nevertheless, the success of these simple and interpretable measures in capturing time reversibility, together with the established presence of these symbolic methods in the literature, motivated a deeper analysis to understand why they perform well.

We identified three high-performing symbolic features that involved counting sequence probabilities in a time series:
the probability of rises in the time series, $p_\text{u}(\bm{x})$ (72\% accuracy), which is a simple normalized index adopted in prior work \cite{porta_temporal_2008}; and
the probability of two consecutive rises (or of two consecutive falls), $p_\text{uu}(\bm{x})$ (and $p_\text{dd}(\bm{x})$) (both have 73\% accuracy).
We also identified five high-performing symbolic indices of irreversibility as the length of consecutive runs of a given symbol, including the excess occurrence of the pattern `uu' (or `dd') relative to a single `u' (or `d'), i.e., $p_\text{uu}(\bm{x}) - p_\text{u}(\bm{x})$ or $p_\text{dd}(\bm{x}) - p_\text{d}(\bm{x})$ (each have 74\% accuracy), as well as the mean length of consecutive rises (or falls) and the difference between the mean lengths of consecutive rises and falls (each have 73\% accuracy).
Figure~\ref{subfig:dist-b}(ii) plots the distribution of $|\Delta p_\text{uu}|$, the best-performing symbolic feature measuring frequencies of specific symbolic motifs within the series, across all reversible and irreversible time series.
As for all the other irreversibility statistics, $|\Delta p_\text{uu}| \approx 0$ for reversible processes, but irreversible processes exhibit substantial deviation from zero.

Similar to the symmetric constructions of generalized autocorrelations (Fig.~\ref{subfig:symm-a}), symbolic sequences that are symmetric with respect to time reversal are invariant under time reversal, i.e., they trivially yield $\Delta f = 0$ by construction (independent of data, see results in Sec.~\ref{subsec:zero_features}).
To clarify what we mean by time-reversal symmetric constructions of symbolic sequences of local rises (u) or falls (d), we note that when a time series is reversed, rises become falls, and vice versa, i.e., $p_\text{u}(\bm{x}) = p_\text{d}(\tilde{\bm{x}})$.
This observation can be formalized using an operator that we call the bitwise `logical negation' operator that maps $\text{u} \mapsto \text{d}$ and $\text{d} \mapsto \text{u}$.
Accordingly, the sequence probabilities of the time-reversed time series $\tilde{\bm x}$ can be expressed in terms of the probability of the logically negated sequence in the original time series $\bm x$: the probability of observing a given sequence in $\tilde{\bm x}$ is equal to the probability of the corresponding bitwise-negated sequence in $\bm x$, read in reverse temporal order (i.e., from right to left).
Symmetric sequences are those for which reading the bitwise-negated sequence in reverse yields the original sequence.
Three examples of time-reversal symmetric sequences---`ud', `udud', and `uudd'---are depicted diagrammatically in Fig.~\ref{subfig:symm-c} with lines indicating rises or falls and the midpoint represented as a red dashed line.
For features constructed with this symmetry, the probability of occurrence in the original and reversed time series is identical, for example $p_{\text{ud}}(\bm x) = p_{\text{ud}}(\tilde{\bm x})$, resulting identically in $\Delta p_\text{ud}(\bm{x}) = p_\text{ud}(\bm{x}) - p_\text{ud}(\tilde{\bm{x}}) = p_\text{ud}(\bm{x}) - p_\text{ud}(\bm{x}) = 0$.

In contrast to these symmetric sequences, for which their time-reversal corresponds to their logical negation, sequences that do not contain this symmetry have the potential to act as a statistical indicator of time reversibility.
Three specific examples of such asymmetric constructions are illustrated in Fig.~\ref{subfig:symm-d}: `uu', `uduu', and `uudu'.
For example, for the `uu' motif, $p_{\text{uu}}(\bm{x})$ (the probability of two successive rises in a time series, 73\% accuracy), we can use the time-reversal property of rises and falls described above to write:
\begin{equation}
    \begin{aligned}
          \Delta p_\text{uu}(\bm{x}) 
          &= p_\text{uu}(\bm{x}) - p_\text{uu}(\tilde{\bm{x}})\,,\\
          &= p_\text{uu}(\bm{x}) - p_\text{dd}(\bm{x})\,.
    \end{aligned}
\end{equation}
This provides an equivalent interpretation of $\Delta p_\text{uu}(\bm{x})$ as a measure of the difference in frequency between two successive rises versus two successive falls in the original time series $\bm{x}$.
Since for a reversible process (with equivalent forward and time-reversed joint distributions, Eq.~\eqref{eq:empirical_rev}), $p_\text{uu}(\bm{x})$ and $p_\text{dd}(\bm{x})$ have equal expectation, an imbalance between $p_\text{uu}(\bm{x})$ and $p_\text{dd}(\bm{x})$ (which yields $\Delta p_\text{uu}(\bm{x}) \neq 0$), can thus act as an statistical indicator of irreversibility.
A similar property applies to all asymmetric sequence probabilities, which are capable of capturing deviations from time-reversibility via imbalances of the frequencies of a sequence (e.g., `uduu') and its logical inversion, including the two examples shown in Fig.~\ref{fig:symmetric-asymmetric}(d), with $\Delta p_\text{uduu}(\bm{x}) = p_\text{uduu}(\bm{x}) - p_\text{ddud}(\bm{x})$ and 
$\Delta p_\text{uudu}(\bm{x}) = p_\text{uudu}(\bm{x}) - p_\text{dudd}(\bm{x})$.

In summary, symbolic sequences provide a simple yet powerful framework for quantifying time-reversal asymmetries in time-series data. 
Similar to generalized autocorrelations, this family of statistics embodies a common symmetry principle while capturing local patterns that may be temporally symmetric (and trivial) or asymmetric (and potentially useful as an index of time-reversal asymmetry).

\subsubsection{Forecasting features}
\label{subsec:forecasting}

The third category of high-performing time-series features we focus on is that involving the simulation of a forecasting algorithm which attempts to predict the future of the time series from its past.
Forecasting algorithms explicitly distinguish the past and future of a time series, and are thus constructed with a directional asymmetry in time.
Despite this, forecasting-based time-series features remain relatively unexplored for this purpose, with the exception of a small number of prior studies (such as the nonlinear prediction model used to detect irreversibility by \citet{stone_detecting_1996}).
The types of high-performing forecasting-based features identified in our data-driven comparison capture various goodness-of-fit metrics of a fitted forecasting model, including the mean absolute error (MAE), and the variance and Gaussianity of the residuals.
The predictive properties of a time series generated by a reversible process remain the same when observed in the forward and reversed directions of time, but there is an asymmetry in irreversible processes that can be detected by forecasting models.

The highest accuracy achieved by a forecasting-based feature was a measure of the Gaussianity of the residuals of a simple local mean forecaster (predicting the subsequent time step using the local mean of the three prior time points (81\% accuracy)).
Other high-performing forecasting-based methods include the goodness of fit of autoregressive (AR) model predictions across various prediction lengths (accuracies in the range 72\%--78\%), with the best-performing feature (78\% accuracy) capturing the goodness of fit of one-step-ahead errors using an AR(2) model.
Other statistics used a range of state-space and GARCH models, and used test statistics from a Kolmogorov--Smirnov test performed on the model residuals (yielding accuracies up to 81\%); or nonlinear prediction error (using a two- or three-dimensional time-delay embedding \cite{michael_small_applied_2005}, achieving accuracies up to 75\%).

Figure~\ref{subfig:dist-b}(iii) illustrates the effectiveness of an example feature for this family, the mean absolute error (MAE) of a one-step-ahead prediction from fitting a second-order autoregressive AR(2) model, in being able to index irreversibility.
As before, and consistent with expectation, we see values tightly concentrated around $|\Delta f| \approx 0$ for reversible processes, and substantial deviations $|\Delta f| > 0$ for irreversible processes.

While useful features show $|\Delta f| > 0$ for some irreversible processes, we were interested to further investigate the sign of the difference $\Delta f$, which indicates whether a time series is more easily predictable forward or backward in time.
Intuitively, a time series generated by some irreversible process will be more predictable in the direction corresponding to the underlying rules of the governing dynamic process, which are formulated forward in time.
This is consistent with the claim of \citet{stone_detecting_1996} that ``theoretically there may be nonlinear systems for which backward predictions might be better than forward predictions, but in practice we have found that such systems, if they exist, must be rare''.
Our results reveal an interesting subtlety to this expectation.
For a sufficiently flexible prediction model, like a nonlinear prediction model (e.g., fitted in a two-dimensional time-delay embedding space \cite{michael_small_applied_2005}), we indeed found a consistent increase in the ability to accurately forecast irreversible processes forward in time (relative to their time-reversed versions).
However, we found counterintuitive results from simpler models, including AR(2) and ARMA(3,1) models, where we found that time-reversal can lead to more accurate predictions for some irreversible processes.
This behavior could be explained by the properties of different types of asymmetries in the $(x_t, x_{t+1})$ phase plots with respect to the $x_t = x_{t+1}$ identity line (which is a hallmark of irreversible dynamics \cite{diks_reversibility_1995}).
One reason may be that the time-reversed series can have properties that better match the assumptions of a simple predictive model.
Relative to the forward-time marginal distributions, $p(x_{t+1}|x_{t})$, time reversal can yield more Gaussian marginals, $p(x_{t}|x_{t+1})$, which can lead to more accurate predictions by simple, strongly parametric models whose assumptions better match the statistical properties of the time-reversed data.

Our results thus highlight an underappreciated class of time-series analysis methods---well-studied algorithms for time-series forecasting---and their ability to capture time reversibility through a difference in predictability in the forward versus time-reversed directions for irreversible processes.
We also report some counterintuitive cases in which some irreversible processes can be more accurately predicted by simple models after time-reversal, due to the reversed series better matching the assumptions of the forecasting model.

\subsection{Statistical signatures of irreversibility are highly process-dependent}
\label{subsec:spectrum_rev}

\begin{figure*}[htbp]
    \centering
    {\includegraphics[width=\linewidth]{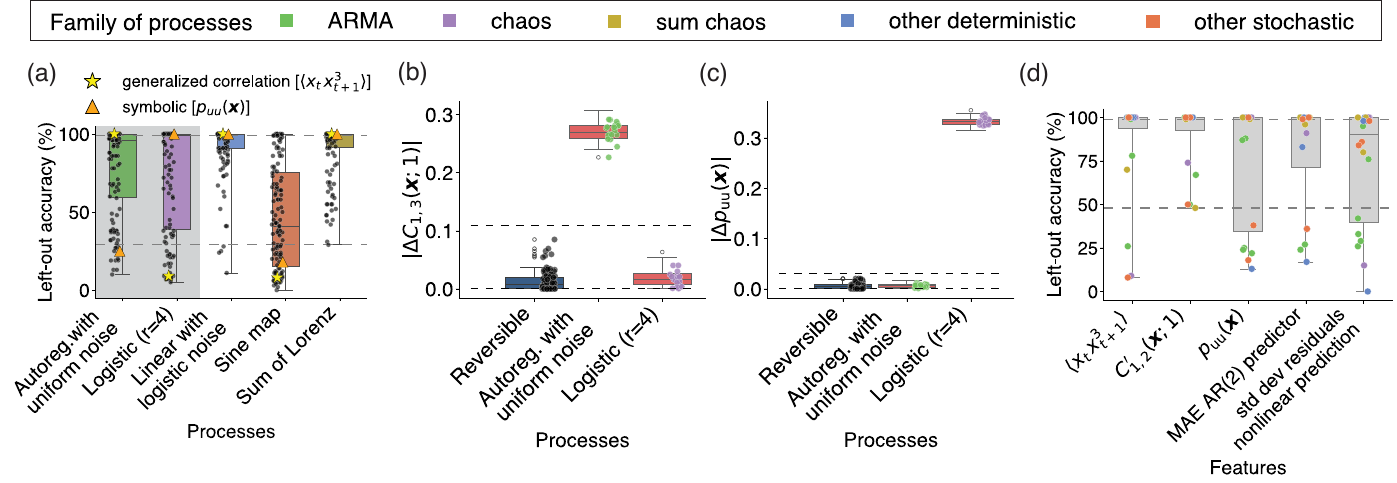}

    \phantomsubcaption\label{subfig:minmax-a}
    \phantomsubcaption\label{subfig:minmax-b}
    \phantomsubcaption\label{subfig:minmax-c}
    \phantomsubcaption\label{subfig:minmax-d}}

    \caption{
    \textbf{All statistical time-series features have strengths and weaknesses at detecting irreversibility across different processes, demonstrating the need to tailor statistical summaries to the specific sources of irreversibility in a given process.}
    \textbf{(a)} Box plot with scatter points showing the left-out accuracy of the 127 top-performing features for five representative irreversible processes: autoregressive with uniform noise distribution (\textsc{AR1\_UNO}), logistic map (with $r = 4$) (\textsc{LOGISTIC\_4}), linear model with logistic map noise (\textsc{LLOG}), noise-driven sine map (\textsc{SINE\_MAP}), and a linear projection of the multidimensional Lorenz system (\textsc{LORENZ\_SUM}).
    Simulation details of the implemented models are in Appendix~\ref{app:models}.
    The generalized autocorrelation feature $\langle x_t\, x_{t+1}^3 \rangle$ is highlighted using a yellow star and the symbolic feature $p_{\text{uu}}(\bm x)$ using an orange triangle.
    For each process, we computed the minimum and maximum left-out accuracies obtained by the set of top-performing features and compared these values across all irreversible processes. We then highlighted the feature with the lowest maximum accuracy across irreversible processes using a dashed line (corresponding to 99\%), and similarly marked the feature with the highest minimum left-out accuracy (29\%).
    We plot the distribution of absolute feature differences $|\Delta f|$ across: (i) all reversible time series; (ii) the autoregressive process with uniform noise (\textsc{AR1\_UNO}); and (iii) the logistic map (with $r = 4$) (\textsc{LOGISTIC\_4}), for two example features: \textbf{(b)} generalized autocorrelation $\langle x_t\, x_{t+1}^3 \rangle$; and 
    \textbf{(c)} the symbolic sequence probability $p_{\text{uu}}$.
    In both plots, the lower solid line denotes the $|\Delta f| = 0$ baseline, while the upper dashed line delineates the maximum of the range of values observed across all time series generated from reversible processes.
    \textbf{(d)} Distribution of left-out accuracy across 2000 time series from the 20 irreversible processes for five representative features: two forms of generalized autocorrelation, namely the fourth-order statistic $\langle x_t\, x_{t+1}^3 \rangle$ and the `generalized linear self-correlation function' $C^\prime_{1,2}(\bm x;1)$ \cite{duarte_queiros_yet_2007}, reported in Eq.~\eqref{eqn:glscf}, using $\alpha = 1$, $\beta = 2$, and $\tau = 1$; the probability of two successive increases (the `uu' pattern) in a time series, $p_{\text{uu}}(\bm x)$; the mean absolute error of 1-step-ahead predictions made by a second-order AR model (labeled as `MAE AR(2) predictor'); and the standard deviation of the residuals from a nonlinear prediction model (labeled as `std dev residuals nonlinear prediction'). 
    }
    \label{fig:minmax}
\end{figure*}

Our results above highlight a range of time-series summary statistics that can index time-reversal asymmetry with high accuracy (up to 89\%) across 35 diverse processes.
However, no feature exhibited perfect separation between time series generated from reversible versus irreversible processes, suggesting that there may not exist a general-purpose real-valued summary statistic that can reliably index time-reversibility.
While average accuracy provides a simple index of a feature's performance, in this section we aimed to understand in more detail the relative strengths and weaknesses of the 127 top-performing features identified above when applied to different types of processes.
This required us to disentangle the relative performance of each feature on each process.
We computed this as the `left-out accuracy', defined for each feature on each irreversible process as the proportion of the 100 time-series realizations correctly classified by the feature (during the testing phase of the 1-NN leave-one-out procedure, cf. Sec.~\ref{sec:methods}).

To understand the differential performance of top-performing features on different processes, we computed the distribution of left-out accuracies across the 127 top-performing features for five irreversible processes, selecting one representative process from each of the five families of processes listed in Table~\ref{tab:processes}: autoregressive with uniform noise distribution (\textsc{AR1\_UNO});
logistic map (with $r = 4$) (\textsc{LOGISTIC\_4});
linear model with noise being a realization of a logistic map (\textsc{LLOG});
noise-driven sine map (\textsc{SINE\_MAP});
and a linear projection of the multidimensional Lorenz system (\textsc{LORENZ\_SUM}).
Results are shown in Fig.~\ref{subfig:minmax-a}.
Different processes exhibit varying levels of difficulty.
The irreversibility of some systems is relatively straightforward to detect (e.g., \textsc{LLOG} and \textsc{LORENZ\_SUM}), where a large number of features achieve high left-out accuracies, whereas other more challenging systems (e.g., \textsc{SINE\_MAP}) show substantially lower accuracies.
Nevertheless, for each of the twenty irreversible processes we tested, there was at least one feature that achieved near-perfect classification accuracy (at least 99\% left-out accuracy, indicated by the upper dashed line in Fig.~\ref{subfig:minmax-a}).
This finding demonstrates that, with access to a sufficiently diverse library of time-reversal sensitive time-series statistics, it is typically possible to formulate a real-valued summary statistic that can accurately index the form through which irreversibility manifests in a given process.
Because the target process was left out during feature selection, the good performance of the statistic result from the association of instances of the selected process with that of other irreversible ones---serving, in this sense, as a meaningful `index of irreversibility'.
In general, however, only a relatively small subset of the (generally high-performing) features can accurately diagnose the irreversibility, in the sense that they preferentially associate instances of a given process in the $\Delta f$ space with other irreversible processes rather than reversible ones. 
Indeed, multiple (of our nominally high-performing) features are unable to distinguish the irreversibility of the process (with a left-out accuracy below chance-level 50\%).
Within the limitations of our comprehensive (but non-exhaustive) tests, this indicates that there exist irreversible processes for which any real-valued `irreversibility index' will fail.
In other words, while there may not be an optimal \textit{a priori} index of irreversibility (as per the various no free lunch theorems \cite{wolpert_no_1997}), our results point to the ability to accurately index irreversibility through tailoring a well-chosen statistic to a given irreversible process of interest.

A demonstrative case study of the strengths and weaknesses of any given statistical indicator of reversibility is illustrated in Fig.~\ref{subfig:minmax-a} for two examples of features analyzed above (see Figs~\ref{subfig:dist-b}(i) and \ref{subfig:dist-b}(ii)): the generalized autocorrelation $\langle x_t\, x_{t+1}^3 \rangle$ (marked as an yellow star) and the probability of the `uu' pattern, $p_{\text{uu}}(\bm x)$ (marked as an orange triangle).
Both features accurately distinguish the irreversibility of the linear model with logistic map noise (\textsc{LLOG}) and the linear projection of the multidimensional Lorenz (\textsc{LORENZ\_SUM}) systems (which are generally easier problems for our statistics), but failed for the noise-driven sine map (\textsc{SINE\_MAP}) system (a more challenging problem for our statistics).
But the two features displayed complementary strengths and weaknesses for the autoregressive with uniform noise (\textsc{AR1\_UNO}) process (for which $\langle x_t\, x_{t+1}^3 \rangle$ is accurate but $p_\text{uu}(\bm x)$ fails) and the logistic map ($r = 4$) (\textsc{LOGISTIC\_4}) process (for which $p_\text{uu}(\bm x)$ is accurate but $\langle x_t\, x_{t+1}^3 \rangle$ fails).
To elucidate this difference, we plotted the absolute difference $|\Delta f|$ for both features across time-series realizations of these two processes (\textsc{AR1\_UNO} and \textsc{LOGISTIC\_4}) in Figs~\ref{subfig:minmax-b} and \ref{subfig:minmax-c}.
These plots demonstrate the ability of $|\Delta C_{1,3}(\bm x;1)|$ to distinguish time series generated from the \textsc{AR1\_UNO} process but not the \textsc{LOGISTIC\_4} process (which have $|\Delta C_{1,3}(\bm x;1)| \approx 0$, and vice-versa for $|\Delta p_\text{uu}(\bm x)|$.

Having characterized the variability of feature performance across individual processes, we now focus on evaluating the relative performance of individual features across the set of simulated processes.
Figure~\ref{subfig:minmax-d} shows the distribution of left-out accuracies for all twenty irreversible processes across five representative top-performing features which capture temporal directionality in the data in diverse yet effective ways: two forms of autocorrelation discussed in Sec.~\ref{subsec:generalized_ac} above, namely $\langle x_t\, x_{t+1}^3\rangle$ and the `generalized linear self-correlation function' \cite{duarte_queiros_yet_2007}, $C^\prime_{1,2}(\bm x;1)$ in Eq.~\ref{eqn:glscf}, the probability of two consecutive rises in the time series $p_{\text{uu}}(\bm x)$, the mean absolute error of 1-step-ahead predictions made by a second-order AR model (denoted as `MAE AR(2) predictor'), and the standard deviation of the residuals given a nonlinear predictor (denoted as `std dev prediction error').
Consistent with their high average accuracies, all features were able to distinguish the irreversibility of the majority of processes, correctly classifying the vast majority of their realizations.
However, each feature exhibits substantial variability in left-out accuracy and, in particular, for each feature there exists at least one process for which it fails to detect irreversibility (using the conservative threshold of having a left-out accuracy below chance level of 50\%).

Taken together, our results reveal a marked degree of inter-process variation in a given feature's ability to detect irreversibility (as assessed empirically here by its ability to preferentially match time series generated from a given irreversible process to that of other irreversible processes).
Each feature encapsulates a specific form of a deviation from time-reversal symmetry (cf. Eq.~\eqref{eq:empirical_rev})---e.g., of a given statistical structure and on a given timescale.
Using a single summary statistic in this way represents a data-efficient simplification of the general problem of measuring deviation in the forward versus time-reversed joint distributions directly (cf. Eq.~\eqref{eq:empirical_rev}), which becomes impractical to assess (particularly on longer lags) from finite time-series data.
But this simplification comes at the cost of imposing a specific structural form of the irreversibility relative to the wide range of statistical structures through which irreversibility can manifest in general (e.g., departures from linearity and Gaussianity on different timescales).
A given feature will perform well on processes that exhibit a given type of time-reversal asymmetry, but relatively poorly on processes that exhibit a different type of asymmetry.
Our findings thus support the view of irreversibility as a multifaceted phenomenon that cannot be fully captured by any single summary statistic in general, but our results suggest a promising way forward in tailoring summary statistics to a given irreversible process.

\section{Discussion}
\label{sec:discussion}
In this work we have presented a unified organization of statistical approaches to detecting irreversibility from time-series data by conducting the broadest comparative study of numerical methods to date, allowing us to characterize and identify connections between a wide array of promising numerical techniques.
By systematically comparing over 6000 statistics across 3500 time series, simulated from a wide range of reversible and irreversible systems, we identified methods that:
(i) recapitulate the effectiveness of established reversibility statistics, including $\langle x_t\,x_{t+1}^3\rangle$ \cite{pomeau_symetrie_1982} and $p_{\text{u}}(\bm x)$ \cite{porta_temporal_2008};
(ii) suggest useful modifications of classical measures, e.g., using absolute values in $\langle |x_t \, x_{t+1}^2| \rangle$ relative to the original test statistic, the TR test statistic $\hat \gamma_{2,1}(1)$ \cite{ramseyj_time_1996};
(iii) demonstrate that statistics originally developed for other purposes can act as powerful metrics of irreversibility, e.g., a formulation of generalized autocorrelation proposed by \citet{duarte_queiros_yet_2007}, or the mean absolute error of 1-step-ahead predictions made by an AR(2) model; and
(iv) highlight novel time-series analysis methods for irreversibility detection that (to our knowledge) have not previously been studied in this context, e.g., a range of forecasting-based statistics, as well as `walker' and extreme-value statistics (Sec.~\ref{subsec:well_features}).
We provide a deeper understanding of our findings by analyzing the algorithmic implementation of effective statistics, illustrated with a novel diagrammatic representation for generalized autocorrelation and symbolic statistics (Fig.~\ref{fig:symmetric-asymmetric}), showing that their asymmetric construction allows them to behave as non-trivial identifiers of irreversibility, consistent with previous studies.
Furthermore, the breadth of comparison conducted here, of both analysis methods and systems, allowed us to demonstrate that there is no general `best statistic' that is effective at detecting irreversibility in all systems; but rather that each statistic captures a specific manifestation of irreversibility, expressed as an asymmetry in some underlying property.
This suggests that the summary statistic could in general be tailored to the specific statistical properties that characterize the time-reversal asymmetry of a given process.
Our results also suggest a range of recommendations.
We argue that future work should aim to benchmark the behavior of different irreversibility statistics across a broad range of systems with different irreversibility signatures (rather than focusing on a small number of hand-picked systems)
And rather than developing new statistics, future research may be more productively focus on developing methods to efficiently tailor statistical indices of irreversibility to a given system of interest.
In summary, our comparative analysis offers a broad view of irreversibility as a multifaceted concept and characterizes the myriad types of time-series analysis methods that can powerfully index it, while also providing insights to guide new directions for its analysis from time-series data.
This will be especially useful for emerging applications of time-irreversibility metrics, including in applications of non-equilibrium statistical thermodynamics, and in inferring and understanding how time-asymmetric structures in time-series data result from underlying dissipative and nonlinear mechanisms in complex physical systems.

The highly comparative data-driven method adopted here compares a large algorithmic library of time-series features derived from diverse types of time-series theory.
This empirical approach involves first simulating time-series data from a range of dynamical mechanisms with known structures (in this case diverse formulations of reversible and irreversible processes), and then systematically searching across an extensive and comprehensive set of candidate features (the \textit{hctsa} feature library \cite{fulcher_hctsa_2017}), to identify and interpret those which are most effective at recovering the underlying time-series structure (irreversibility).
Prior highly comparative analyses following this general methodological template have identified statistical indicators of self-affine time series \cite{fulcher_highly_2013} and the distance to criticality from short noisy time series \cite{harris_tracking_2024}.
This empirical approach is complementary to a theory-driven approach, whereby new statistical methods are derived from the theory of irreversible processes (e.g., \citet{diks_reversibility_1995}).
By drawing on a wide array of existing statistical theory, here we are able to provide a unified interdisciplinary perspective on the common conceptual formulations of time-series structure that are informative of time-reversibility.
We also distinguish our data-driven approach, as a wide comparison of time-series analysis methods, from other data-driven machine-learning approaches; the connection to time-series theory is essential in enabling us to connect patterns in the data to an interpretable theory for understanding and guiding future advancements.

Our analysis highlights new relationships between diverse formulations of irreversibility, bridging methods developed across different scientific fields.
We highlight important similarities between generalized autocorrelation statistics (Sec.~\ref{subsec:generalized_ac}) and symbolic sequences derived from time-series increments (Sec.~\ref{subsec:symbolic}): in both cases, symmetric structures are insensitive to reversibility, while asymmetric statistics provide sensitive real-valued indicators of irreversibility.
In the case of generalized autocorrelations, asymmetry arises through the choice of weights or temporal lags, while for symbolic features it emerges from asymmetric patterns in successive rises and falls; in both cases, trivial statistics correspond to time-symmetric local patterns, and deviations from this symmetry give rise to families of statistics with relatively high predictive performance.
This finding aligns with the observation of \citet{steinberg_time_1986} for continuous signals, namely that ``appropriate measures being those that depend on the direction of the \textit{time arrow}''.
The variety of constructions identified here resonates with the considerations of \citet{pomeau_symetrie_1982}, who emphasized the potentially infinite formulations of asymmetric autocorrelation functions as effective proxies for detecting irreversibility.
The distinction between symmetric and asymmetric constructions is illustrated in Fig.~\ref{fig:symmetric-asymmetric}, where we introduced a diagrammatic representation that makes the underlying temporal symmetries visually clear.
We also found connections between methodologies adopted within the same family.
For example, we identified a linear relationship between two generalized forms of autocorrelation (under ergodicity and stationarity) that originated from distinct fields: the measure of \citet{schreiber_discrimination_1997} (the $t^\mathrm{rev}(\tau)$ statistic) developed in nonlinear dynamics, and the `TR test' statistic of \citet{ramseyj_time_1996} (the $\hat{\!\gamma}_{2,1}(\tau)$ statistic) developed in finance, in Eq.~\eqref{eq:similar_sch_rr}.
Among the symbolic features, our result that the proportion of rises (or falls), as well as its extensions to rise/fall patterns (e.g., the proportion of two consecutive rises or falls), aligns with methods adopted in prior works, e.g., the percentage of positive variations (that is the percentage of rises in a time series) introduced to detect temporal asymmetries in heart rate variability series \cite{porta_time_2006}.
Additionally, through our broad inclusion of statistics across the time-series analysis literature we were able to identify novel time-series statistics for capturing time-reversibility that have not previously (to our knowledge) been used for this purpose.
We flag a broad range of forecasting approaches, whether linear or nonlinear, that could be adapted to this problem through differences in predictability between models applied in the forward versus backward directions.
These statistics performed strongly relative to the limited prior work on this topic, yielding time series that are generally more predictable in the forward direction when using a nonlinear predictor, consistent with the findings of \citet{stone_detecting_1996}, while also highlighting counterintuitive cases in which the time-reversed series is more easily predictable when using linear models.
Complementing these findings, statistics derived from simulating dynamical processes on the time series proved to be among the most effective.
These features offer potentially novel contributions to the time-series literature, spanning from statistics of simulated `walkers' to measures inspired by extreme-value theory \cite{altmann_reactions_2006}.

Across the wide range of simulated processes, no individual tested time-series feature could detect irreversibility across all analyzed processes.
As discussed in Sec.~\ref{sec:intro}, full information on reversibility is encoded in the time-reversal symmetry of the joint distribution $p({\bm  x}_t^{(\tau)})$ which, for reasonable values of the temporal lag $\tau$, is impractical to estimate from finite data; any statistic that reduces this information to a single value thus discards information.
Consequently, each real-valued time-series summary statistic developed with the purpose of detecting time reversibility captures a specific form of temporal asymmetry (tied to a specific time-series property and typically on a predefined timescale, or set of timescales).
The limited information captured by any individual time-series statistic helps explain why previous studies, often focusing on single approaches, sometimes reached contrasting conclusions about the same processes \cite{zanin_algorithmic_2021}.
Indeed, the breadth of our comparisons confirm that a method effective for one process may fail for another, highlighting the intrinsic limitation of any single-feature approach.
At the same time, we show that (given the leave-one-process-out nearest-neighbor matching heuristic used here) the large algorithmic library of time-series statistics contained in \textit{hctsa} \cite{fulcher_hctsa_2017} is sufficiently comprehensive to identify an accurate statistic for capturing time irreversibility for all simulated processes.

Recognizing the strengths and weaknesses of each time-series feature lays the groundwork for future investigations into explicit hypothesis tests, facilitated by the development of an appropriate null distribution \cite{diks_reversibility_1995}.
This would extend the 1-NN approach used here as a practical heuristic index, under the assumption that the set of reversible and irreversible processes is sufficiently comprehensive.
While this heuristic was adequate for our current purpose, which primarily focuses on comparative analysis, future work could aim to establish a formal testing framework that more precisely characterizes statistical deviations from reversibility.
For example, an empirical treatment could involve defining a broad set of reversible processes as the null distribution for reversibility, and then quantifying the deviation for a given time series (and test statistic) using a permutation test.
While we have briefly confirmed relatively similar behavior of key irreversibility indices as a function of time-series length (as shown in Fig.~\ref{fig:appC-robustness}), future work could more comprehensively evaluate the generalization of our results to short, noisy time series (where different features may exhibit different levels of robustness to time-series length and additive noise).
Our analysis here was also restricted to univariate time series; future work could investigate extending this approach to multivariate data, building on approaches such as cross-correlation analysis widely used in neuroscience \cite{deco_insideout_2022}, noting that a comprehensive library of pairwise statistics has recently been developed \cite{cliff_unifying_2023}.
Finally, while our range of systems analyzed here is comprehensive, it is not exhaustive, and our specific quantitative and qualitative results depend on these choices.
Future work could therefore evaluate a broader range of processes, including those containing more complex timescales \cite{zanin_algorithmic_2025}.

\begin{acknowledgments}
T.D.N. acknowledges support from the Australian Research Council (DP240101295).
B.D.F. acknowledges support from the Australian Research Council (FT240100418).
\end{acknowledgments}

\appendix

\section{Simulation details of implemented models}
\label{app:models}
Here we present the simulated processes in greater detail, including the governing equations and specifying the parameter choices made in this work.
In the summary Table~\ref{tab:processes} we indicated with a `D' the discrete-time and with a `C' the continuous-time processes to emphasize that distinct simulation strategies are required for continuous-time systems which include greater attention to initial condition selection, downsampling procedures, and integration parameters, as discussed in detail in Sec.~\ref{subsec:time_series_generation}.

\paragraph*{Noise.}
Time-series realizations of independent and identically distributed (i.i.d.) noise processes were drawn from both a Gaussian $x_t \sim \mathcal{N}(0,1)$ (labeled \textsc{GNO}) and a uniform distribution $x_t \sim \mathcal{U}(-0.5,0.5)$ (labeled \textsc{UNO}).
Three colored noise processes were generated using Zhivomirov's algorithm\cite{zhivomirov_method_2018}, implemented in MATLAB. 
As mentioned in Sec.~\ref{subsec:model_systems}, we simulated realizations of pink noise, characterized by a power spectral density (PSD) proportional to $ 1/f$ (labeled \textsc{PINK}), where $f$ is frequency; red noise with a PSD proportional to $1/f^2$ (labeled \textsc{RED}); and violet noise with a PSD proportional to $f^2$ (labeled \textsc{VIOLET}).

\paragraph*{Autoregressive models.}
We simulated a first-order autoregressive process, AR(1), given by
\begin{equation}
    x_t = a\,x_{t-1} + \varepsilon_t\,,
    \label{eq:AR1}
\end{equation}
where $a = 0.5$ and $\varepsilon_t \sim \mathcal{N}(0,1)$ is i.i.d. Gaussian noise (labeled \textsc{AR1\_GNO}).
Additionally, we considered the static nonlinear transformation of another AR(1) process given by Eq.~\eqref{eq:AR1} with $a = 0.6$, as $y_t = \tanh^2(x_t)$ (labeled \textsc{STAR}). 
We adopted the same or closely related parameter choices reported in \citet{diks_reversibility_1995}.

Next, we simulated autoregressive processes with various forms of non-Gaussianities and nonlinearities.
Linear non-Gaussian stochastic processes simulated here include:
\begin{itemize}
    \item[i.] the AR(1) in Eq.~\eqref{eq:AR1} with noise terms sampled from a uniform distribution, $\varepsilon_t \sim \mathcal{U}(-0.5, 0.5)$ (labeled \textsc{AR1\_UNO}). We decided to maintain the same parameter choice ($a=0.5$) for consistency with \textsc{AR1\_GNO}, in order to compare two processes with the same deterministic component and, therefore, isolate the noise contribution;
    
    \item[ii.] an autoregressive ARMA(1,1) process:
    \begin{equation}
          x_t = 0.6\,x_{t-1} + \epsilon_t + 0.4\, \epsilon_{t-1}\,,
    \end{equation}
    with noise sampled from a uniform distribution $\varepsilon_t \sim \mathcal{U}(-0.5, 0.5)$ (labeled \textsc{ARMA11\_UNO}); 

    \item[iii.] an autoregressive processes with Gamma noise distribution, defined as:
    \begin{equation}
        x_t = 0.3\, x_{t-3} - 0.2\, x_{t-2} + 0.1\, x_{t-1} +\epsilon_t\,,
    \end{equation}
    with manually specified parameters and noise sampled from gamma distribution, $\epsilon_t \sim \Gamma(1,0.3)$ (labeled \textsc{AR3\_GAMMA}). 
    The initial condition was drawn from a uniform distribution over $[0,1)$. 
\end{itemize}

We also considered three nonlinear autoregressive processes.
The first form of nonlinearity was introduced through Self-Exciting Threshold Autoregressive (SETAR) models, which switch between $k$ different autoregressive regimes based on the past $d$ values of the time series \cite{tong_threshold_1980}.
A SETAR model is denoted SETAR$(k; d,p_1,...,p_k)$ where $k$ represents the number of regimes, $d$ is the delay parameter that controls the switch between regimes and $p_1,...,p_k$ are the orders of the autoregressive models within each of the $k$ regimes.
We simulated two variants of this process, namely the SETAR$(2;1,1,1)$ model defined as \cite{mittnik_higher-order_1999} (labeled \textsc{SETAR1}):
\begin{equation}
    x_t=\begin{cases}
        -0.9\, x_{t-1} + \epsilon_t & \text{if }x_{t-1} \geq 1\,,\\
        -0.4\, x_{t-1} + \epsilon_t & \text{if } x_{t-1} < 1\,,
        \end{cases}
\end{equation}
and the following specific SETAR$(2;2,2,2)$ process introduced by \citet{martinez_detection_2018}:
\begin{equation}
    x_t =\begin{cases}
        0.62 + 1.25\, x_{t-1} -0.43\, x_{t-2} + 0.0381\, \varepsilon_t,~~ \text{if } x_{t-2}\leq 3.25\,,\\
        2.25 + 1.52\, x_{t-1} -1.24\, x_{t-2} + 0.0626\, \varepsilon_t,~~ \text{if } x_{t-2} > 3.25\,,
        \end{cases}
\end{equation}
where $\varepsilon_t \sim \mathcal{N}(0,1)$ are i.i.d. (labeled \textsc{SETAR2}).
For the initial conditions, the first point of the \textsc{SETAR1} process and the first two points of the \textsc{SETAR2} process were sampled from a standard normal distribution.
The second form of nonlinear process is defined by the following equation:
\begin{equation}
    x_{t+1} = 0.5\, x_t-0.3\, x_{t-1} + 0.1\, y_{t-1} + 0.1\, x_{t-1}^2 + 0.4\, y_t^2 + 0.0025\, \eta_t\,,
\end{equation}
where 
\begin{equation}
     y_t = \sin(4\pi t) + \sin(6\pi t) + 0.0025\, \xi_t\,,
\end{equation}
which is driven by both Laplacian $\eta_t\sim \text{Laplace}(0,1)$ (sharper peak and heavier tails comped to Gaussian distribution) and bimodal Gaussian $\xi_t\sim 0.5~\mathcal{N}(0.63,1) + 0.5~\mathcal{N}(-0.63,1)$ noise (labeled \textsc{NAR2}).
The first two points of both the $x$ and $y$ components were set to zero. 

\paragraph*{Deterministic chaotic maps.}
We simulated the conservative (measure-preserving) Arnold's Cat map \cite{arnold_ergodic_1968} (labeled \textsc{ARNOLD}), defined by:
\begin{equation}
    \begin{aligned}
        x_{t+1}&= x_t + y_t \,\text{mod}~1\,,\\
        y_{t+1}&= x_t + 2 y_t\,\text{mod}~1\,,
    \end{aligned}
\end{equation}
and the Chirikov standard map \cite{roberts_chaos_1992} (labeled \textsc{CHIRIKOV}):
\begin{equation}
    \begin{aligned}
     \theta_{t+1} &= \theta_t +p_t + \frac{K}{2\pi}\,\sin{(2\pi\theta_t)} ~~~\text{mod}~1\,,\\
    p_{t+1} &= \theta_{t+1} - \theta_t~~~ \text{mod}~1\,,        
    \end{aligned}
\end{equation}
with parameter $K = 0.971635$.
For our analysis we considered the $x$-component of the Arnold's Cat map and the $p$-component of the Chirikov standard map. 
As representatives of dissipative chaotic maps, we examined the $x$-component of the well-studied standard H\'enon map \cite{henon_two_1976} ($a = 1.4$ and $b = 0.3$) (labeled \textsc{HEN}):
\begin{equation}
    \begin{aligned}
    x_{t+1} &= 1 - a\,x_t^2 + y_t\,, \\
    y_{t+1} &= b\,x_t\,,
\end{aligned}
\end{equation}
as well as two additional one-dimensional chaotic systems.
The first one is a quadratic map \cite{sprott_simplest_1997} given by the equation $x_{t+1} = 1 - a\,x_t^2$, with choice of the parameter $a=1.8$ (labeled \textsc{QUAD}).
The second, the logistic map \cite{may_simple_1976}, is given by $x_{t+1} = r\, x_t\,(1 - x_t)$, which we simulated both in fully chaotic regime ($r = 4$) (labeled \textsc{LOGISTIC\_4}) and in the period-3 window ($r = 3.8284$), where it exhibits intermittency (labeled \textsc{LOGISTIC\_38}).
For all chaotic maps, initial conditions were randomly sampled from a uniform distribution over the unit interval, $[0,1)$.

\paragraph*{Sum of deterministic chaotic maps/flows.}
We constructed time series by combining two independent realizations of the $x$-component (one realization taken in forward time and another in reverse-time order, constructed by flipping the second in time) of two discrete-time systems: (i) the H\'enon map (labeled \textsc{HENR\_DIVERSE}); and (ii) quadratic map (labeled \textsc{QUAD\_RSUM}).
As a special case for the H\'enon map, we also simulated and included a process obtained by summing a forward-time realization with its time-reversed counterpart, generated by flipping the same time series (labeled \textsc{HENR\_SAME}).
In addition, we summed pairs of independent forward realizations of the $x$-component of the standard H\'enon map (labeled \textsc{HEN\_SUM}).

We included continuous-time processes in our analysis, considering a linear projection $x + y + z$ of two multidimensional systems: the Lorenz and R\"ossler systems.
We simulated realizations of the Lorenz system \cite{lorenz_deterministic_1963} with equations ($\sigma = 10$, $\rho = 28$, $\beta = 8/3$):
\begin{equation}
    \begin{aligned}
    \frac{dx}{dt} &= \sigma\, (y - x)\,,\\
    \frac{dy}{dt} &= x\,(\rho - z) - y\,, \\
    \frac{dz}{dt} &= x\,y - \beta\, z\,,
    \label{eq:lorenz}
\end{aligned}
\end{equation}
with initial conditions sampled from uniform distributions $x_0, y_0 \sim \mathcal U(-8,8)$ and $z_0 \sim \mathcal U(0,10)$ (labeled \textsc{LORENZ\_SUM}).
The second chaotic flow analyzed is the R\"ossler system \cite{rossler_equation_1976} given by ($a = 0.2$, $b = 0.2$ and $c = 5.7$):
\begin{equation}
    \begin{aligned}
    \frac{dx}{dt} &= -y - z\,,\\
    \frac{dy}{dt} &= x + a\,y\,, \\
    \frac{dz}{dt} &= b + z\,(x-c)\,,
\end{aligned}
\end{equation}
for which the initial conditions were sampled from the uniform distribution $x_0, y_0 \sim \mathcal U(-8,8)$ and $z_0 = 0$ (labeled \textsc{ROSSLER\_SUM}]). 

\paragraph*{Other deterministic models.}
We generated time series from two discrete-time deterministic models:
\begin{itemize}
    \item[i.] the linear model with deterministic noise source, $x_{t+1} = a\,x_{t} + \varepsilon_t$, where $a = 0.5$ and $\varepsilon_t$ is a realization of a logistic model $\varepsilon_t = 4\,\varepsilon_{t-1}\,(1-\varepsilon_{t-1})$ (labeled \textsc{LLOG}) that we refer to as `logistic noise';
    \item[ii] the process $x_t = a\,x_{t-1}~\text{mod}~1$, where $a = \sqrt{2}$ is the parameter that rules the reversible nature of the process (labeled \textsc{MOD1}).
\end{itemize}
Additionally, we included two continuous-time systems described by second-order differential equations:
\begin{itemize}
    \item[i.] a linearized oscillator with dynamics described by the equation:
\begin{equation}
    \frac{dx^2}{dt^2} + x = 0\,,
\end{equation}
with initial conditions sampled from a uniform distribution $x_0, y_0 \sim \mathcal U(-2,2)$ (labeled \textsc{OSCILLATOR});

    \item[ii.] the Van der Pol oscillator \cite{van_der_pol_jun_lxxxviii_1926} specified by:
\begin{equation}
    \frac{d^2x}{dt^2} - \mu\,(1 - x^2)\, \frac{dx}{dt} + x = 0\,,
    \label{eq:vanDerPol}
\end{equation}
where $\mu = 1$, to include effects due to non-linear damping and initial conditions sampled uniformly at random from the $[-2,2)$ interval (labeled \textsc{VDP}).
\end{itemize}
We simulated dynamics Mackey--Glass equations, dynamics characterized by temporal delays \cite{mackey_oscillation_1977}.
Specifically, we analyzed the following delay differential equation:
\begin{equation}
    \frac{dx}{dt} = \frac{\beta_0\ x(t - \tau)}{1 + x(t-\tau)^n} - \gamma \ x(t)\,,
\end{equation}
where we set $\beta_0 = 0.2$, $\gamma = 0.1$, $n = 10$, and the delay $\tau = 17$ (labeled \textsc{MG17}).
The initial condition was sampled from a uniform distribution, $x_0 \sim \mathcal U(0,1.5)$.

\paragraph*{Other stochastic processes.}
As a discrete-time system, we simulated a noise-driven sine map:
\begin{equation}
    x_{t+1} = \mu \sin{(x_t)} + y_t \eta_t\,,
\end{equation}
where $\mu = 2.4$ and $y_t$ is Bernoulli random variable that is $1$ with probability $0.01$ and $0$ otherwise and $\eta_t \sim \mathcal{U}(-2,2)$ \cite{freitas_failure_2009} (labeled \textsc{SINE\_MAP}).

In continuous time, we simulated the Ornstein--Uhlenbeck process given by:
\begin{equation}
    dx = -\theta\, x~dt +\sigma\,dW\,,
\end{equation}
with $\theta = 0.8$, $\sigma = 0.3$, $W$ a Wiener process and initial condition $x_0 = 0$ (labeled \textsc{OU}).

We also included bounded random walks (BRW), which are random walks adjusted by a function that biases the random walk downward towards a neighborhood of the stationary mean $\tau$ when the process deviates excessively from it \cite{nicolau_stationary_2002}. 
The introduction of this adjustment function guarantees the global stationary of the generated time series, limiting the trend behavior typical of random walks.
The continuous-time version of BRW simulated here is given by:
\begin{equation}
    dx = e^k(e^{-\alpha_1(x - \tau)}- e^{\alpha_2(x - \tau)} )dt + e^{\sigma/2+\beta/2x_t^2}dW\,,
\end{equation}
with parameters $k = -2$, $\alpha_1 = \alpha_2 = 2$, $\sigma = 4$, $\beta = 0.1$,  $x_0 = 0$ and $W$ a standard Wiener process (labeled \textsc{BRW}).
Finally, we included stochastic variants of the Lorenz and Van der Pol systems. The stochastic Lorenz system (labeled \textsc{STOCH\_LORENZ\_SUM}) was obtained by adding independent  Wiener processes, $\eta_x,\eta_y$ and $\eta_z$ to the deterministic dynamics pf each coordinate in Eq.~\eqref{eq:lorenz}, and the Van der Pol oscillator (labeled \textsc{STOCH\_VDP}) was generated by adding a standard Wiener process $W$ to Eq.~\eqref{eq:vanDerPol}.
Both systems were initialized as in the deterministic case, and the noise intensity was set by scaling the Wiener processes by a factor of $1.5$.

\section{Robustness of time-series statistics for reversibility to the time-series length}
\label{app:robustness}

\renewcommand{\thefigure}{B.\arabic{figure}} \setcounter{figure}{0}
 
\begin{figure*}[htbp]
    \centering
    \includegraphics[width=0.75\linewidth]{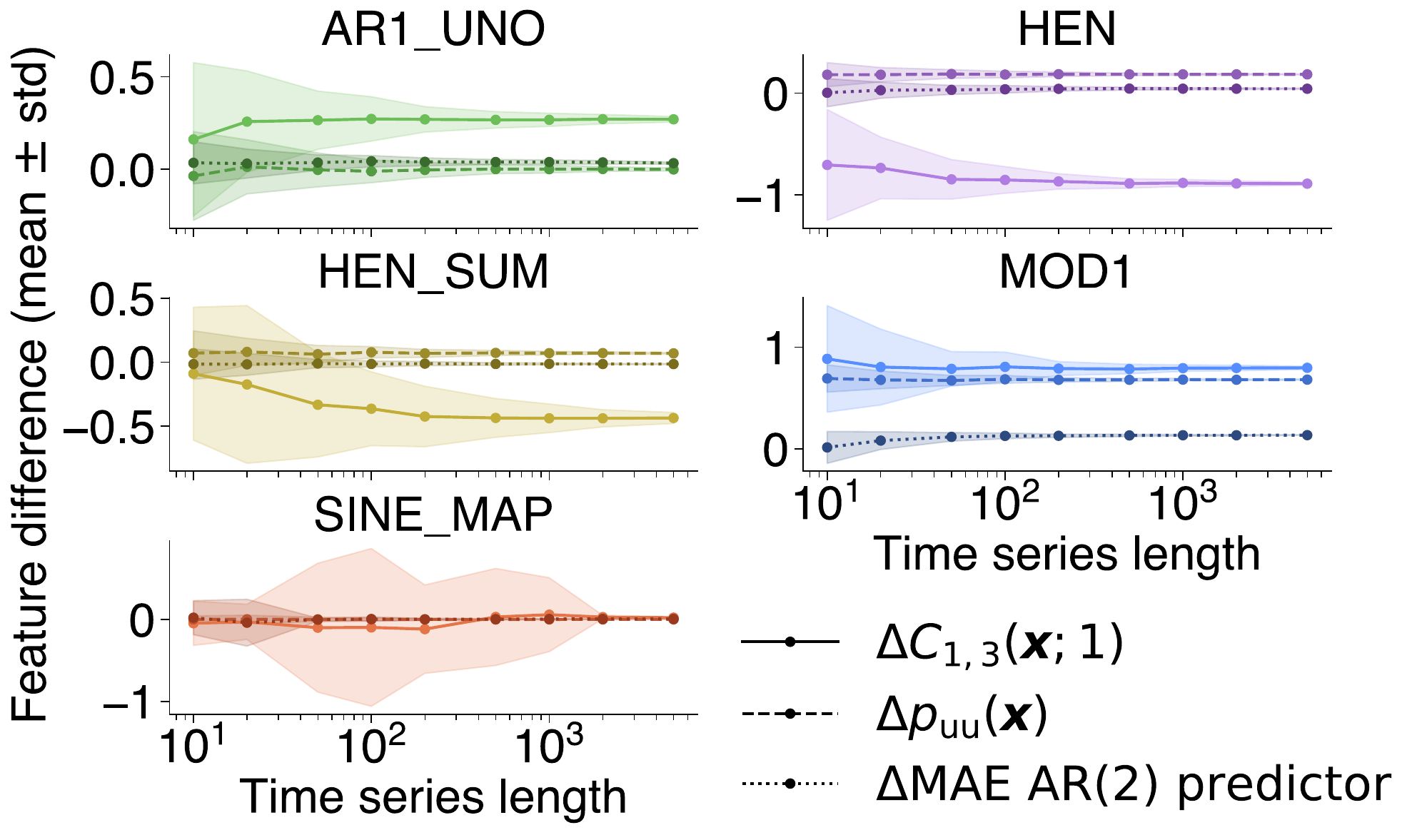}
    \caption{\textbf{Time-series features $\langle x_t^3\, x_{t+1}\rangle$, $p_{\text{uu}}(\bm x)$ and mean absolute error (MAE) of an AR(2) predictor as a function of the time-series length for a representative process from each family.}
    We show the mean and standard deviation of the feature difference of the generalized autocorrelation $\langle x_t^3\, x_{t+1}\rangle$ (light colors, solid line) and the probability of two successive rises $p_{\text{uu}}(\bm x)$ (medium colors, dashed line) and MAE of an AR(2) predictor (dark colors, dotted line) for time series of lengths $T$= [10, 20, 50, 100, 200, 500, 1000, 2000, 5000] ($x$-axis, logarithmic scale), computed over 100 realizations of each process.
    Each panel reports a representative process per family, with colors indicating process families (see Table~\ref{tab:processes}).}
    \label{fig:appC-robustness}
\end{figure*}

The main analysis in this work was carried out considering 5000-sample time series.
To assess the robustness of the selected features with respect to time-series length, we computed the mean and standard deviation of the feature differences $\Delta f$ for a subset of the top-performing statistics discussed in Sec.~\ref{subsec:well_features}, across nine different lengths of discrete-time processes: $T \in \{10, 20, 50, 100, 200, 500, 1000, 2000, 5000\}$.
Results are depicted in Fig.~\ref{fig:appC-robustness} for the generalized autocorrelation $\langle x_t^3\, x_{t-1}\rangle $ (light colors, solid line), the symbolic $p_{\text{uu}}(\bm x)$ (medium colors, dashed line) statistics, and the mean absolute error (MAE) of an AR(2) predictor (dark colors, dotted line) for a representative process per family (see Table~\ref{tab:processes}).
Overall, we observe a larger variability in the generalized autocorrelation compared to the symbolic and the forecasting-based statistics, which exhibits a relatively flat trend for $T>50$. 
For shorter time series, errors in estimating the statistics increase, leading to higher variability in the results, particularly for the generalized autocorrelation.

\section{Analysis of the time-reversal transformation of the linear autocorrelation}
\label{app:symm_ac}
As a simple demonstration of the invariance of the two-point linear autocorrelation $\langle x_t \, x_{t+\tau}\rangle$ to time reversal when computed on a finite time series of length $T$, we consider the temporal translation transformation $t \mapsto T-t-\tau+1$ (it reduces to the continuous reversal $t \mapsto -t$ in the limit $T \to \infty$), which preserves the number of time points in the average, $t\in[1,T-\tau]$.
Using this transformation, the two-point linear autocorrelation computed on the reversed time series $\tilde{\bm x}$, where each term is defined as $\tilde x_t=x_{T-t+1}$, can be expressed in terms of the original time series $\bm x$
\[
\langle \tilde x_{t}\, \tilde x_{t+\tau}\rangle = \langle  x_{t+\tau}\, x_t\rangle
\]
which is equivalent to $\langle x_t\, x_{t+\tau} \rangle$. 
Consequently, the corresponding feature difference is identically zero, irrespective of the reversibility of the process that generated the time series.

\section{Absolute difference of asymmetric two-point generalized autocorrelation}
\label{app:asymm_2ac}
Here we compute explicitly the absolute difference for the general formulation of a two-point generalized autocorrelation statistic $C_{\alpha,\beta}(\bm x;\tau)=\langle x_t^\alpha\, x_{t+\tau}^\beta \rangle$ using the shifted time-reversal transformation $t\mapsto T-t-\tau+1 \equiv s$ which preserves the number of points in the time average ($t,s\in [1, T-1]$).
We can express the two-point autocorrelation $C_{\alpha,\beta}(\tilde{\bm x};\tau)$ in terms of the original time series $\bm x$ as:
\begin{equation}
    \begin{aligned}
        C_{\alpha,\beta}(\tilde{\bm x};\tau) &=\langle \tilde{x}_t^\alpha\, \tilde{x}_{t+\tau}^\beta \rangle = \frac{1}{T-\tau -1} \sum_{t=1}^{T-\tau} x_{T-t+1}^\alpha \, x_{T-t-\tau +1}^\beta \,, \\
        &=\frac{1}{T-\tau -1} \sum_{s=1}^{T-\tau} x_{s+\tau}^\alpha \, x_s^\beta\,= \langle x_{t+\tau}^\alpha \, x_t^\beta\rangle \,.
    \end{aligned}
\end{equation}
Consequently, the absolute difference can be written as
\begin{equation}
    \begin{aligned}
         |\Delta C_{\alpha,\beta}(\bm{x};\tau)| &= |C_{\alpha,\beta}(\bm{x};\tau) - C_{\alpha,\beta}(\tilde{\bm{x}};\tau)|\,, \\
         &=|\langle x_t^\alpha \, x_{t+\tau}^\beta\rangle - \langle x_{t+\tau}^\alpha \, x_t^\beta\rangle|\,.
    \end{aligned} 
\end{equation}

\section{Absolute difference of asymmetric three-point generalized autocorrelation}
\label{app:asymm_3ac}
Here we compute explicitly the absolute difference for the general case of a three-point generalized autocorrelation statistic, $|\Delta C_{\alpha,\beta,\gamma}(\bm x;\tau_1,\tau_2)|$.
The calculation is similar to the reasoning that brought to Eq.~\eqref{eq:C_twoPoint} but here we emphasize the dependence on both the choice of the temporal lags $\tau_1,\tau_2 $ and of the exponents $\alpha,\beta,\gamma$ to build symmetric constructions. 
We consider the general form of a three-point generalized autocorrelation
\begin{equation}
    C_{\alpha,\beta,\gamma}(\bm{x};\tau_1,\tau_2) = \langle x_t^\alpha \, x_{t+\tau_1}^\beta \, x_{t+\tau_2}^\gamma\rangle\,,
    \label{eq:general_form_Ac}
\end{equation}
for positive, integer exponents $\alpha$, $\beta$ and $\gamma$, and time-lags $\tau_1, \tau_2$, such that $\tau_2 \neq 2 \tau_1$ if $\alpha = \gamma$.
Since we have two time-lags $\tau_1$ and $\tau_2$, an appropriate shifting transformation is $t \mapsto T-t-\tau_\mathrm{max} + 1$, where $\tau_\mathrm{max} = \text{max}(\tau_1, \tau_2)$.
By writing $C_{\alpha,\beta,\gamma}(\tilde{\bm x};\tau_1,\tau_2)$ in terms of $\bm x$, as for the two-point generalized autocorrelation, the absolute feature difference gets
\begin{equation}
    \begin{aligned}
        &|\Delta C_{\alpha, \beta, \gamma} (\bm x; \tau_1, \tau_2)|= |C_{\alpha,\beta,\gamma}(\bm{x};\tau_1, \tau_2) - C_{\alpha,\beta,\gamma}(\tilde{\bm{x}};\tau_1, \tau_2)|\\&= |\langle x_t ^\alpha \, x_{t+\tau_1}^\beta \, x_{t+\tau_2}^\gamma \rangle - \langle x_{t+\tau_\mathrm{max}}^\alpha  \, x_{t+(\tau_\mathrm{max} - \tau_1)}^\beta \, x_{t + (\tau_\mathrm{max} - \tau_2)}^\gamma \rangle|\,.
    \end{aligned}
\end{equation}
This formulation extends Eq.~\eqref{eq:C_twoPoint} and can itself be generalized to longer temporal scales in two ways: (i) by considering higher-order products, which incorporate longer histories, or (ii) by evaluating pairs or triplets of observations separated by larger temporal lags.
Therefore, the test statistic $|\Delta C_{\alpha, \beta, \gamma} (\bm x; \tau_1, \tau_2)|$ captures deviations from reversibility by quantifying the asymmetry in the data structure arising from the combined effects of unequal weights and differing temporal lags between data-points.

\renewcommand{\thefigure}{A.\arabic{figure}} \setcounter{figure}{0}

\bibliography{biblio}

\end{document}